\documentclass[english,reprint,longbibliography]{revtex4-2}
\usepackage[T1]{fontenc}
\usepackage[latin9]{inputenc}
\usepackage{color}
\usepackage{babel}
\usepackage{mathtools}
\usepackage{amsmath}
\usepackage{amssymb}
\usepackage{graphicx}
\usepackage{xargs}[2008/03/08]
\usepackage[pdfusetitle,
 bookmarks=true,bookmarksnumbered=true,bookmarksopen=true,bookmarksopenlevel=1,
 breaklinks=false,pdfborder={0 0 1},backref=false,colorlinks=true]
 {hyperref}
\hypersetup{
 pdfborderstyle=}

\makeatletter
\usepackage{color}
\usepackage{babel}
\usepackage{xargs}[2008/03/08]
\usepackage[justification=raggedright]{caption}
\usepackage{ragged2e}
\DeclareCaptionJustification{justified}{\justifying}
\ifdefined\showcaptionsetup
 \PassOptionsToPackage{caption=false}{subfig}
\fi
\usepackage{subfig}
\makeatother

\begin{document}
\title{Efficient adaptive control strategy for multi-parameter quantum metrology
in two-dimensional systems}
\author{Qifei Wei$^{1}$}
\author{Shengshi Pang$^{1,2}$}
\email{pangshsh@mail.sysu.edu.cn}

\affiliation{$^{1}$School of Physics, Sun Yat-sen University, Guangzhou, Guangdong
510275, China}
\affiliation{$^{2}$Hefei National Laboratory, Hefei 230088, China}
\begin{abstract}
Quantum metrology leverages quantum resources such as entanglement
and squeezing to enhance parameter estimation precision beyond classical
limits. While optimal quantum control strategies can assist to reach
or even surpass the Heisenberg limit, their practical implementation
often requires the knowledge of the parameters to be estimated, necessitating
adaptive control methods with feedback. Such adaptive control methods
have been considered in single-parameter quantum metrology, but not
much in multi-parameter quantum metrology so far. In this work, we
bridge this gap by proposing an efficient adaptive control strategy
for multi-parameter quantum metrology in two-dimensional systems.
By eliminating the trade-offs among optimal measurements, initial
states, and control Hamiltonians through a system extension scheme,
we derive an explicit relation between the estimator variance and
evolution time. Through a reparameterization technique, the optimization
of evolution times in adaptive iterations are obtained, and a recursive
relation is established to characterize the precision improvement
across the iterations. The proposed strategy achieves the optimal
performance up to an overall factor of constant order with only a
few iterations and demonstrates strong robustness against deviations
in the errors of control parameters at individual iterations. Further
analysis shows the effectiveness of this strategy for Hamiltonians
with arbitrary parameter dependence. This work provides a practical
approach for multi-parameter quantum metrology with adaptive Hamiltonian
control in realistic scenarios.
\end{abstract}
\maketitle
\global\long\def\rhoa{\rho_{\boldsymbol{\alpha}}}%
 
\global\long\def\cov{C}%
 
\global\long\def\tn{n}%
 \newcommandx\pd[1][usedefault, addprefix=\global, 1=]{\partial_{#1}}%
 \newcommandx\unitdag[1][usedefault, addprefix=\global, 1=\boldsymbol{\alpha}]{U_{#1}^{\dagger}}%
 
\global\long\def\unita{U_{\boldsymbol{\alpha}}}%
 \newcommandx\unitt[2][usedefault, addprefix=\global, 1=\boldsymbol{\alpha}, 2=t]{U_{#1}\left(#2\right)}%
 \newcommandx\gen[1][usedefault, addprefix=\global, 1=]{h_{#1}}%
 \newcommandx\ha[1][usedefault, addprefix=\global, 1=\boldsymbol{\alpha}]{H_{#1}}%
 \newcommandx\hainit[2][usedefault, addprefix=\global, 1=\boldsymbol{\alpha}, 2=\mathrm{init}]{H_{#1}^{\left(#2\right)}}%
 
\global\long\def\hcont{H_{\mathrm{c}}}%
 
\global\long\def\iA{I_{\mathrm{A}}}%
 \newcommandx\gent[2][usedefault, addprefix=\global, 1=, 2=t]{h_{#1}\left(#2\right)}%
 
\global\long\def\deltae{\delta E}%
 
\global\long\def\deltaes{\delta E^{2}}%
 \newcommandx\deltaesk[1][usedefault, addprefix=\global, 1=]{\delta E_{#1}^{2}}%
 \newcommandx\fa[2][usedefault, addprefix=\global, 1=, 2=]{f_{#1}\left(#2\right)}%
 \newcommandx\hcontkth[1][usedefault, addprefix=\global, 1=]{H_{\mathrm{c},#1}}%
 \newcommandx\fav[1][usedefault, addprefix=\global, 1=]{\boldsymbol{f}\left(#1\right)}%
 \newcommandx\deltaak[1][usedefault, addprefix=\global, 1=]{\boldsymbol{\delta}\boldsymbol{\alpha}_{#1}}%
 \newcommandx\deltabk[1][usedefault, addprefix=\global, 1=]{\boldsymbol{\delta}\boldsymbol{\beta}_{#1}}%
 \newcommandx\deltabki[2][usedefault, addprefix=\global, 1=, 2=]{\delta\beta_{#1,#2}}%
 \newcommandx\deltaeskave[1][usedefault, addprefix=\global, 1=]{\left\langle \delta E_{#1}^{2}\right\rangle }%
 \newcommandx\covk[1][usedefault, addprefix=\global, 1=]{C_{#1}}%
 \newcommandx\covkb[1][usedefault, addprefix=\global, 1=]{C_{#1}^{\left(\boldsymbol{\beta}\right)}}%
 \newcommandx\covka[2][usedefault, addprefix=\global, 1=, 2=\boldsymbol{\alpha}]{C_{#1}^{\left(#2\right)}}%
 \newcommandx\prodz[1][usedefault, addprefix=\global, 1=0]{g_{#1}}%
 
\global\long\def\prodd{g}%
 \newcommandx\toptk[2][usedefault, addprefix=\global, 1=\mathrm{opt}, 2=]{t_{#1,#2}}%
 \newcommandx\prodzs[2][usedefault, addprefix=\global, 1=0, 2=2]{g_{#1}^{#2}}%
 \newcommandx\gain[1][usedefault, addprefix=\global, 1=]{G\left(#1\right)}%
 \newcommandx\dev[1][usedefault, addprefix=\global, 1=k]{D_{#1}}%
 \newcommandx\tnk[1][usedefault, addprefix=\global, 1=]{n_{#1}}%
 \newcommandx\vsum[1][usedefault, addprefix=\global, 1=]{V_{#1}}%
 \newcommandx\tk[1][usedefault, addprefix=\global, 1=]{t_{#1}}%
 \newcommandx\reso[1][usedefault, addprefix=\global, 1=]{T_{#1}}%
 \newcommandx\resoopt[1][usedefault, addprefix=\global, 1=]{T_{\mathrm{opt},#1}}%
 \newcommandx\resotot[1][usedefault, addprefix=\global, 1=]{T_{\mathrm{tot},#1}}%
 
\global\long\def\psiz{\left|\psi_{0}\right\rangle }%
 
\global\long\def\psizct{\left\langle \psi_{0}\right|}%
 \newcommandx\prob[1][usedefault, addprefix=\global, 1=]{p_{#1}}%
 \newcommandx\lambdai[1][usedefault, addprefix=\global, 1=]{\lambda_{#1}}%
 \newcommandx\bv[1][usedefault, addprefix=\global, 1=]{l_{#1}}%
 \newcommandx\bvdtf[1][usedefault, addprefix=\global, 1=]{0_{#1}}%
 \newcommandx\bvdts[1][usedefault, addprefix=\global, 1=]{1_{#1}}%
 \newcommandx\genm[1][usedefault, addprefix=\global, 1=]{h_{#1}^{\left(\mathrm{M}\right)}}%
 
\global\long\def\deltaf{\delta F}%
 
\global\long\def\gap{\kappa}%
 \newcommandx\realvei[1][usedefault, addprefix=\global, 1=]{\mu_{#1}}%
 \newcommandx\imvei[1][usedefault, addprefix=\global, 1=]{\nu_{#1}}%
 \newcommandx\ts[1][usedefault, addprefix=\global, 1=]{t_{\mathrm{s},#1}}%
 \newcommandx\eigen[1][usedefault, addprefix=\global, 1=]{E_{#1}}%
 
\global\long\def\tnoc{n_{\mathrm{oc}}}%
 
\global\long\def\tkoc{t_{\mathrm{oc}}}%
 \newcommandx\vsumt[1][usedefault, addprefix=\global, 1=]{\widetilde{V}_{#1}}%

Precision measurement plays a fundamental role across various disciplines
of science and technology. Quantum metrology \citep{giovannetti2006quantum,paris2009quantum,giovannetti2011advances},
rooted in the principles of quantum mechanics and statistical inference,
exploits nonclassical resources such as entanglement and squeezing
to realize estimation of parameters in quantum dynamics with high
precision. This technique has been widely applied in atomic interferometers
\citep{jacobson1995quantum}, atomic clocks \citep{andre2004stability,borregaard2013nearheisenberglimited,kessler2014heisenberglimited},
gravitational wave detection \citep{schnabel2010quantum,theligoscientificcollaboration2011agravitational}
and so on. Over the past decades, quantum metrology has seen rapid
advancement in both theoretical innovation and experimental breakthroughs.

Theoretically, entangled probes evolving in parallel under parameter-dependent
dynamics can achieve the Heisenberg limit \citep{giovannetti2006quantum,pang2014quantum}.
Alternatively, a sequential strategy---where a single probe evolves
under the parameter-dependent dynamic and adaptive control---also
achieves the Heisenberg limit and offers advantages when entanglement
is difficult to generate or maintain \citep{yuan2015optimal,yuan2016sequential,pang2017optimal}.
Quantum metrology considers both single-parameter estimation \citep{braunstein1994statistical,kolodynski2013efficient,demkowicz-dobrzanski2014usingentanglement}
and multi-parameter \citep{fujiwara2001estimation,humphreys2013quantum,yuan2016sequential,suzuki2016explicit,baumgratz2016quantum,szczykulska2016multiparameter,proctor2018multiparameter,altenburg2018multiparameter,yang2019optimal,albarelli2019evaluating,carollo2019onquantumness,sidhu2021tightbounds,xia2023towardincompatible,hou2020minimal,albarelli2020aperspective,suzuki2020quantum,suzuki2020nuisance,belliardo2021incompatibility,katariya2021rldfisher,lu2021incorporating,gorecki2022multiparameter,albarelli2022probeincompatibility}
estimation. While single-parameter estimation has been well understood,
the multi-parameter quantum metrology poses additional challenges
due to the incompatibility of the optimal measurements, initial states,
and control strategies with respect to different parameters \citep{carollo2019onquantumness,sidhu2021tightbounds,xia2023towardincompatible}.
Besides, environmental noise is inevitable in realistic scenarios,
and significant progress has been made in addressing quantum metrology
for open systems, both in exploring estimation precision limits \citep{fujiwara2008afibre,escher2011general,demkowicz-dobrzanski2012theelusive,kolodynski2013efficient}
and developing noise-resilient strategies \citep{sekatski2016dynamical,dong2016reviving,zhou2018achieving,gorecki2020optimal,chen2018achieving}.

Experimentally, quantum metrology has been implemented on a variety
of physical systems, e.g., photonic systems \citep{leibfried2004towardheisenberglimited,nagata2007beating,resch2007timereversal,chen2018achieving,chen2018heisenbergscaling,hou2021zerotradeoff,markiewicz2021simultaneous,xia2023towardincompatible,yin2023experimental},
nuclear magnetic resonance \citep{lu2020observing,zhai2023controlenhanced},
superconducting circuits \citep{naghiloo2017achieving}, etc. These
experiments have realized key theoretical breakthroughs, such as attaining
the Heisenberg limit \citep{bollinger1996optimal}, improving the
efficacy by control-enhanced strategies \citep{zhai2023controlenhanced},
full estimation of magnetic fields \citep{hou2020minimal,hou2021zerotradeoff},
and mitigating the incompatibility of multi-parameter estimation \citep{xia2023towardincompatible},
etc.

In quantum metrology, quantum control serves as a powerful tool to
boost the estimation precision. In noiseless scenarios, Hamiltonian
control has shown the capability of increasing the precision to the
Heisenberg limit and even beyond \citep{pang2017optimal}. The optimal
control strategies have been well established for single-parameter
estimation, including both time-independent and time-dependent Hamiltonians
\citep{yuan2015optimal,pang2017optimal}. For multi-parameter quantum
metrology, significant progress has also been made in two-dimensional
systems \citep{yuan2016sequential,hu2024control} where the system
extension scheme eliminates the trade-offs completely, but the realizability
of optimal quantum control in practice remains much less explored.

The optimal control Hamiltonian usually relies on the knowledge of
the unknown parameters to be estimated. This necessitates the use
of adaptive control strategies to iteratively refine the control Hamiltonian
based on the estimated values of the parameters from previous measurements.
Although preliminary work has addressed adaptive control for the single-parameter
estimation \citep{pang2017optimal}, efficient adaptive strategies
for multi-parameter scenarios remain largely an open problem. Moreover,
while one can certainly enhance the estimation precision by an increasing
number of iterations and trials in each iteration, quantum resources
are limited for any quantum protocol. So it is crucial to design an
efficient adaptive control strategy that enables rapid convergence
to the optimal Hamiltonian control with given resources.

In this work, we bridge this gap by introducing an efficient adaptive
control strategy tailored for multi-parameter quantum metrology. Considering
the feasibility of analytical computation, we focus our research on
two-dimensional systems, similar as most studies of multi-parameter
quantum metrology with quantum control have pursued \citep{fujiwara2001estimation,yuan2016sequential,baumgratz2016quantum,hou2021zerotradeoff,yang2022multiparameter,hu2024control},
but the analysis can be effective for general quantum systems. We
analyze the time dependence of the estimation variances of unknown
parameters, and elucidate the mechanism underlying the Hamiltonian
control strategy that can achieve the Heisenberg limit. By integrating
a system extension scheme with iterative feedback control and leveraging
the reparameterization technique, we design an efficient adaptive
control strategy for estimating the three orthogonal components of
a qubit Hamiltonian in the Pauli basis, which can eliminate the trade-offs
among measurements, initial states, and control strategies and achieves
the optimal precision up to an overall factor with only a few iterations
while maintaining the robustness against deviations in the errors
of control parameters. Furthermore, we prove the general applicability
of our approach to Hamiltonians with arbitrary parameter dependence,
making it a practical tool for quantum metrology in realistic experimental
scenarios.

\section*{Results}

\textbf{Quantum multi-parameter estimation theory.} In quantum single-parameter
estimation, the quantum Cramer-Rao bound tells that the variance of
an estimator is bounded by the inverse of the quantum Fisher information,
as the quantum Fisher information characterizes the sensitivity of
a parameter-dependent quantum state to the variations in the parameter
\citep{wootters1981statistical,sidhu2020geometric}. For multi-parameter
estimation, the quantum Fisher information can be extended mathematically
to the quantum Fisher information matrix \citep{helstrom1967minimum,liu2019quantum}.
However, the quantum Cramér-Rao bound based on symmetric logarithmic
derivatives is not always attainable due to the potential incompatibility
between the optimal measurements for different parameters \citep{carollo2019onquantumness,albarelli2020aperspective,sidhu2021tightbounds,xia2023towardincompatible,hou2020minimal},
unless specific conditions are satisfied, e.g. the weak commutativity
in the asymptotic limit of collective measurement on an unlimited
number of systems \citep{matsumoto2002anew,ragy2016compatibility,demkowicz-dobrzanski2020multiparameter}
or a more strict condition when the number of accessible systems is
finite \citep{yang2019optimal,chen2022incompatibility}. Therefore,
estimating multiple unknown parameters in a quantum state is a challenging
problem.

Suppose a quantum state $\rhoa$ depends on $q$ unknown parameters
denoted in a vector form $\boldsymbol{\alpha}=\left(\alpha_{1},\alpha_{2},\ldots,\alpha_{q}\right)$.
The estimation precision of the unknown parameters is characterized
by the covariance matrix $\cov$, which is bounded by the quantum
Fisher information matrix $F$,
\begin{equation}
\cov\geq\left(\tn F\right)^{-1},\label{eq:qme-1}
\end{equation}
where ``$\geq$'' represents the matrix semi-definite positivity
and $\tn$ refers to the number of trials. The entries of quantum
Fisher information matrix are given by
\begin{equation}
F_{ij}=\frac{1}{2}\mathrm{Tr}\left(\rhoa\left\{ L_{i},L_{j}\right\} \right),\label{eq:qme-2}
\end{equation}
where $L_{i}$ is a symmetric logarithmic derivative defined by
\begin{equation}
2\pd[i]\rho_{\boldsymbol{\alpha}}=\rho_{\boldsymbol{\alpha}}L_{i}+L_{i}\rho_{\boldsymbol{\alpha}},\label{eq:qme-3}
\end{equation}
where $\pd[i]$ is the abbreviation of $\pd[\alpha_{i}]$ for simplicity.
In reality, the unknown parameters in a quantum state are usually
encoded by physical processes. If the physical process is a unitary
evolution $\unita$ and the initial state of the quantum system is
$\left|\psi_{0}\right\rangle $, the entries of quantum Fisher information
matrix are given by
\begin{equation}
F_{ij}=4\left(\frac{1}{2}\left\langle \left\{ \gen[i],\gen[j]\right\} \right\rangle -\left\langle \gen[i]\right\rangle \left\langle \gen[j]\right\rangle \right),\label{eq:qme-4}
\end{equation}
where $\gen[i]\coloneqq-i\left(\pd[i]\unitdag\right)\unita$ is the
generator of the infinitesimal translation of $\unita$ with respect
to the $i$-th parameter $\alpha_{i}$, $\left\{ \ ,\ \right\} $
denotes the anti-commutator, and $\left\langle \cdot\right\rangle =\left\langle \psi_{0}\right|\cdot\left|\psi_{0}\right\rangle $.

To address the potential incompatibility issue between the optimal
measurements for different parameters, a real symmetric matrix $W$
can be introduced to assign weights to different parameters and define
a weighted mean precision $\mathrm{Tr}\left(W\cov\right)$ as the
overall benchmark for the performance of estimation. The lower bound
of this weighted mean precision can be derived from the quantum Cramér-Rao
bound,
\begin{equation}
S\left(W\right)=\frac{1}{\tn}\mathrm{Tr}\left(WF^{-1}\right).\label{eq:qme-5}
\end{equation}
The weighted mean precision can be optimized and attain the Holevo
bound in the asymptotic limit of the number of trials if collective
measurements on multiple quantum systems are allowed \citep{albarelli2019evaluating,demkowicz-dobrzanski2020multiparameter,sidhu2021tightbounds}.
But the Holevo bound is usually hard to be solved explicitly, as it
still involves a complex matrix optimization problem. Nevertheless,
when a weak commutativity condition is satisfied, i.e., for any two
parameters $\alpha_{i}$ and $\alpha_{j}$,
\begin{equation}
\mathrm{Tr}\left(\rhoa\left[L_{i},L_{j}\right]\right)=0,\label{eq:qme-6}
\end{equation}
the quantum Cramér-Rao bound coincides with the Holevo bound and can
therefore be attained \citep{matsumoto2002anew,ragy2016compatibility}.
For the estimation of parameters $\boldsymbol{\alpha}$ in a unitary
operator $U_{\boldsymbol{\alpha}}$, if the initial state of the system
is $\left|\psi_{0}\right\rangle $, the weak commutativity condition
can be further simplified as
\begin{equation}
\mathrm{Im}\left\langle \gen[i]\gen[j]\right\rangle =0,\label{eq:qme-7}
\end{equation}
and independent measurements on individual systems are sufficient
to achieve the quantum Cramér-Rao bound in this case \citep{fujiwara2001estimation,matsumoto2002anew}.

\textbf{Multi-parameter quantum metrology in two-dimensional systems.}
Quantum metrology can generally be decomposed to four steps: preparation
of the initial states, parameter-dependent evolution, measurements
on the final states, and post-processing of the measurement results
to extract the parameters. The estimation precision can be improved
by initial state optimization, feedback control, and measurement optimization
at the first three steps and by using proper estimation strategies
at the final step. In multi-parameter quantum metrology, the incompatibility
issue lies in several aspects: in addition to the measurement incompatibility,
the optimal initial states and optimal feedback controls for different
parameters can be incompatible as well.

\begin{figure}
\includegraphics[scale=0.5]{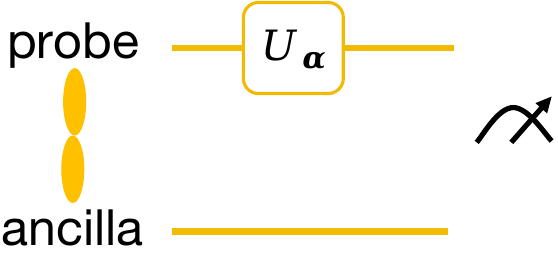}

\caption{\textbf{System extension scheme.} An ancilla with the same dimension
as the probe is introduced, with the unitary evolution $\protect\unita$
acts only on the probe. The initial state can be any quantum state
of the joint system, and measurements are performed on the joint system.}\label{fig:system_extension}
\end{figure}

System extension scheme has been widely used in quantum metrology,
for instance, to establish upper bounds on the quantum Fisher information
in noisy environments \citep{escher2011general,kolodynski2013efficient}
and to eliminate the aforementioned incompatibilities in multi-parameter
quantum estimation \citep{fujiwara2001estimation,yuan2016sequential}.
Fig. \ref{fig:system_extension} shows the system extension scheme,
where a probe and an ancilla are coupled. The unitary evolution $\unitt=\exp\left(-i\ha t\right)$
governed by the parameter-dependent Hamiltonian $\ha$ acts on the
probe only.

When the joint system is initialized in a maximally entangled state,
the weak commutativity condition is satisfied, eliminating the measurement
tradeoff \citep{fujiwara2001estimation,yuan2016sequential} (see Supplementary
Note 1). In two-dimensional systems, this configuration enables optimal
estimation of all the parameters via projective measurements along
the Bell basis, addressing the initial-state trade-off issue \citep{fujiwara2001estimation,yuan2016sequential}
(see Supplementary Note 2). Furthermore, when the initial Hamiltonian
$\hainit$ is independent of time, feedback control using the reverse
of $\hainit$ obtains the optimal estimation for all the parameters
and achieves the Heisenberg limit (see Supplementary Note 3), thereby
removing the control trade-off \citep{yuan2016sequential}.

However, the optimal control Hamiltonian depends on the true values
of the unknown parameters, so an adaptive control is generally required
to update the value of the parameters in the control Hamiltonian iteratively,
so that the control Hamiltonian can approach the optimum progressively.
Such an adaptive feedback control scheme has been studied for the
single-parameter quantum metrology \citet{pang2017optimal}, but has
not received much investigation in multi-parameter quantum metrology.
The dependence of the control Hamiltonian on the parameter estimation
precisions from the previous rounds at each iteration makes it challenging
to evaluate the overall performance of the adaptive procedure and
design efficient iterative feedback strategies, even for two-dimensional
systems.

\textbf{Variance-time relation.} With all trade-offs eliminated, we
analyze the time dependence of the variances of the parameters to
be estimated, offering guidance for the design of efficient adaptive
control strategies.

In adaptive Hamiltonian control, the total Hamiltonian $\ha$ comprises
the initial Hamiltonian $\hainit$ and a control Hamiltonian $\hcont$.
In a two-dimensional system with an ancilla, the joint Hamiltonian
is $\ha\otimes\iA$, where $\iA$ denotes the two-dimensional identity
operator on the ancilla. The joint evolution of the probe and the
ancilla is $\unitt\otimes\iA$, where $\unitt=\exp\left(-i\ha t\right)$,
and the generator of the infinitesimal translation of $\unitt\otimes\iA$
with respect to the parameter $\alpha_{i}$ is $\gent[i]\otimes\iA$,
where
\begin{equation}
\gent[i]=-i\left(\pd[i]\unitt^{\dagger}\right)\unitt.\label{eq:vr-1}
\end{equation}

We choose a maximally entangled state,
\begin{equation}
\psiz=\left(\left|\bvdtf[\mathrm{P}]\bvdtf[\mathrm{A}]\right\rangle +\left|\bvdts[\mathrm{P}]\bvdts[\mathrm{A}]\right\rangle \right)/\sqrt{2},\label{eq:vr-2}
\end{equation}
as the initial state, where $\left\{ \left|\bvdtf[\mathrm{P}]\right\rangle ,\left|\bvdts[\mathrm{P}]\right\rangle \right\} $
and $\left\{ \left|\bvdtf[\mathrm{A}]\right\rangle ,\left|\bvdts[\mathrm{A}]\right\rangle \right\} $
are sets of complete orthogonal basis for the probe and the ancilla,
respectively. According to Eq.~\eqref{eq:qme-4}, the entries of
quantum Fisher information matrix can be obtained as
\begin{equation}
F_{ij}\left(t\right)=2\mathrm{Tr}\left(\gent[i]\gent[j]\right)-\mathrm{Tr}\left(\gent[i]\right)\mathrm{Tr}\left(\gent[j]\right).\label{eq:vr-3}
\end{equation}

To make the time dependence of the quantum Fisher information matrix
explicit for the design of adaptive control strategy, we first analyze
the generator $\gent[i]$. By applying an integral formula for the
derivative of an operator exponential,
\begin{equation}
\frac{\partial\mathrm{e}^{M_{\alpha}t}}{\partial\alpha}=\int_{0}^{t}\mathrm{e}^{M_{\alpha}\left(t-\tau\right)}\frac{\partial M_{\alpha}}{\partial\alpha}\mathrm{e}^{M_{\alpha}\tau}d\tau,\label{eq:vr-4}
\end{equation}
we obtain
\begin{equation}
\gent[i]=\int_{0}^{t}\mathrm{e}^{i\ha\tau}\left(\partial_{i}\ha\right)\mathrm{e}^{-i\ha\tau}d\tau.\label{eq:vr-5}
\end{equation}
Through the spectral decomposition of the Hamiltonian of the probe,
$\ha=\eigen[0]\left|\eigen[0]\right\rangle \left\langle \eigen[0]\right|+\eigen[1]\left|\eigen[1]\right\rangle \left\langle \eigen[1]\right|$
\citep{wilcox1967exponential,pang2014quantum}, we obtain
\begin{equation}
\begin{array}{rl}
F_{ij}\left(t\right)= & \sum_{l=0}^{1}t^{2}\left[\left(\pd[i]E_{l}\right)\left(\pd[j]E_{l}\right)-\left(\pd[i]E_{l}\right)\left(\pd[j]E_{1-l}\right)\right]\\
 & -8\sin^{2}\left(\frac{\deltae t}{2}\right)\left\langle E_{l}\right|\left.\pd[i]E_{1-l}\right\rangle \left\langle E_{1-l}\right|\left.\pd[j]E_{l}\right\rangle ,
\end{array}\label{eq:vr-6}
\end{equation}
where $\deltae$ is the energy gap of the probe Hamiltonian, $\deltae=\eigen[0]-\eigen[1]$.
The complete derivation is provided in Supplementary Note 4. This
equation characterizes the time dependence of quantum Fisher information
matrix: the major term grows quadratically with time, while the other
oscillates with time.

The estimation precision of parameters is characterized by the estimator
variances, which are bounded by the diagonal elements of the inverse
of the quantum Fisher information matrix and the number of trials.
As the system is two-dimensional, we assume $\boldsymbol{\alpha}=(\alpha_{1},\alpha_{2},\alpha_{3})$.
The results for the estimator variances are given in Supplementary
Note 5, where a detailed analysis is provided. Since the estimator
variances for different parameters are symmetric, we only present
the estimator variance for the first parameter as an example. The
estimator variance for the first parameter $\alpha_{1}$ can be obtained
as
\begin{equation}
\left\langle \delta^{2}\widehat{\alpha}_{1}\right\rangle =\frac{1}{n}\frac{\csc^{2}\left(\frac{1}{2}\deltae t\right)t^{2}\xi_{1}+\xi_{2}}{t^{2}\xi_{3}},\label{eq:vr-7}
\end{equation}
where
\begin{equation}
\begin{array}{rl}
\xi_{1}= & \left(\realvei[2]\pd[3]\deltae-\realvei[3]\pd[2]\deltae\right)^{2}+\left(\imvei[2]\pd[3]\deltae-\imvei[3]\pd[2]\deltae\right)^{2},\\
\xi_{2}= & 16\left(\realvei[3]\imvei[2]-\realvei[2]\imvei[3]\right)^{2},\\
\xi_{3}= & 16\left[\realvei[1]\left(\imvei[3]\pd[2]\deltae-\imvei[2]\pd[3]\deltae\right)+\realvei[2]\left(\imvei[1]\pd[3]\deltae-\imvei[3]\pd[1]\deltae\right)\right.\\
 & \left.+\realvei[3]\left(\imvei[2]\pd[1]\deltae-\imvei[1]\pd[2]\deltae\right)\right]^{2},
\end{array}\label{eq:vr-8}
\end{equation}
and $\realvei[i]$ and $\imvei[i]$ are given by
\begin{equation}
\begin{array}{rl}
\realvei[i]= & \mathrm{Re}\left(\left\langle \eigen[0]\right|\left.\partial_{i}\eigen[1]\right\rangle \right),\\
\imvei[i]= & \mathrm{Im}\left(\left\langle \eigen[0]\right|\left.\partial_{i}\eigen[1]\right\rangle \right).
\end{array}\label{eq:vr-9}
\end{equation}

\begin{figure}
\subfloat[\label{fig:variance_time-a}]{\includegraphics[scale=0.35]{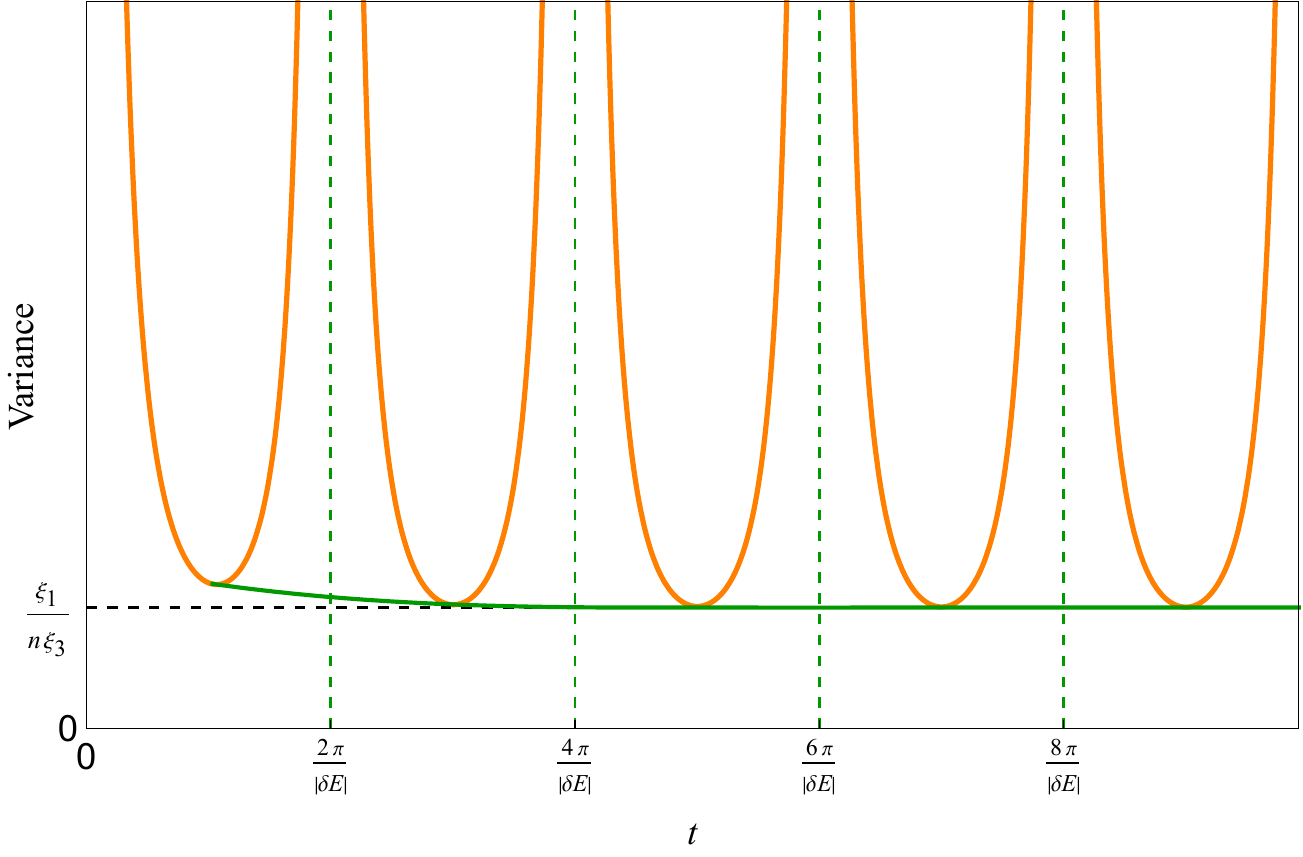}

}

\subfloat[\label{fig:variance_time-b}]{\includegraphics[scale=0.35]{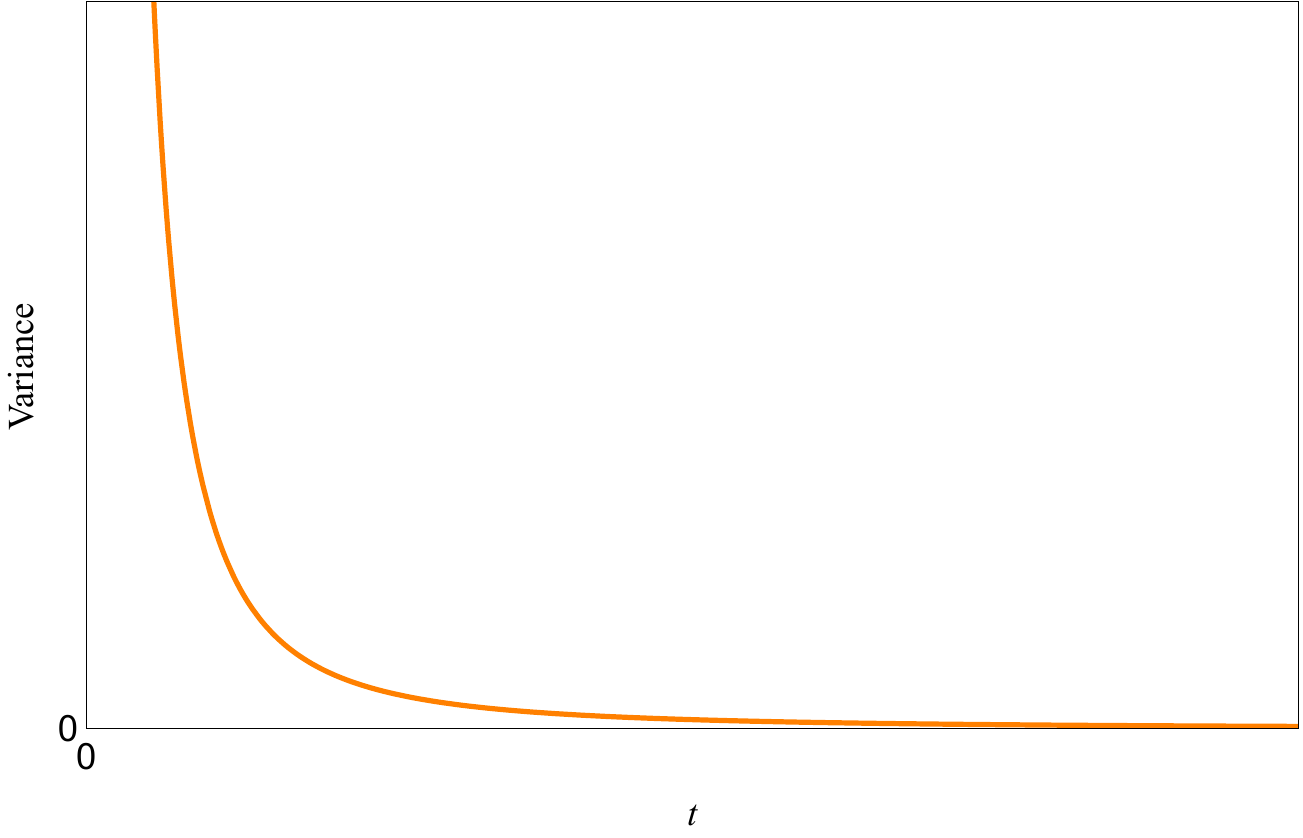}

}

\caption{\textbf{Relation between estimation variance and evolution time.}
The estimation variance of $\widehat{\alpha}_{1}$ exhibits two characteristic
time scalings. Fig. (a) shows the time scaling of variance for $\xi_{1}\protect\neq0$,
depicted by the orange curve. For $t\ll1/\left|\protect\deltae\right|$,
the variance decays quadratically with time. As $t$ increases, the
variance oscillates with time and diverges at integer multiples of
$2\pi/\left|\protect\deltae\right|$, with the asymptotes plotted
by the green dashed lines. The lower envelope of the variance, depicted
by the green solid line, decays and rapidly converges to $\xi_{1}/n\xi_{3}$.
Fig. (b) illustrates the time scaling of estimation variance for $\xi_{1}=0$,
where the variance decays quadratically with time and reaches the
Heisenberg scaling.}\label{fig:variance_time}
\end{figure}

The variance exhibits only two characteristic time scalings, as shown
in Fig. \ref{fig:variance_time}. Fig. \ref{fig:variance_time-a}
shows the time scaling of variance for the case with $\xi_{1}\neq0$:
when $t\ll1/\left|\deltae\right|$, $\csc^{2}\left(\deltae t/2\right)\approx4/\left(\deltae t\right)^{2}$,
hence the estimation variance for the first parameter decays quadratically
with time, while for a longer evolution time $t$, the variance oscillates
at a frequency of $\left|\deltae\right|/2\pi$, with its lower envelope
decaying and rapidly converging to $\xi_{1}/n\xi_{3}$. Fig. \ref{fig:variance_time-b}
shows the time scaling of variance for $\xi_{1}=0$, where the variance
achieves the Heisenberg scaling.

As aforementioned, the optimal control Hamiltonian is $\hcont=-\hainit$,
but its dependence on the unknown parameters poses a practical challenge.
The adaptive control strategy provides a solution to this challenge:
an initial estimation of the unknown parameters is performed without
quantum control to yield a rough approximate value for the parameters
used in the control Hamiltonian; in the following rounds, the control
Hamiltonian is implemented with the estimated values of the unknown
parameters from the previous rounds, resulting in improved precisions.
This process is iterated for multiple rounds. As $\hcont$ approaches
$-\hainit$, the total Hamiltonian $\ha$ as well as $\deltaes$ converges
toward zero. Based on the preceding results, the decrease of $\deltaes$
extends the evolution time $t$ satisfying $t\ll1/\left|\deltae\right|$
during which the estimator variances decay quadratically with time.
Consequently, the estimation precision asymptotically approaches the
Heisenberg limit (see Supplementary Note 3). Therefore, in order to
achieve the Heisenberg-limited time scaling for an evolution time
as long as possible, our objective is to design a strategy that reduces
$\deltaes$ rapidly with a given total evolution time.

The dependence of the Hamiltonian $H_{\boldsymbol{\alpha}}$ on $\boldsymbol{\alpha}$
can be arbitrary in general. To simplify the following study, we reparameterize
the Hamiltonian as
\begin{equation}
\hainit[\boldsymbol{\beta}]=\beta_{1}\sigma_{x}+\beta_{2}\sigma_{y}+\beta_{3}\sigma_{z},\label{eq:vr-10}
\end{equation}
with $\beta_{1}$, $\beta_{2}$, and $\beta_{3}$ being the parameters
to be estimated which are transformed from $\alpha_{1}$, $\alpha_{2}$,
$\alpha_{3}$ and the superscript $({\rm init})$ denotes that it
is the initial Hamiltonian without any control. The necessity of reparameterization
is shown in Supplementary Note 6. But we will also show the validity
of our results for arbitrary parameter dependence of the Hamiltonian
later. The control Hamiltonian at the $\left(k+1\right)$-th iteration
is denoted by
\begin{equation}
\hcontkth[k+1]=-\hat{\boldsymbol{\beta}}_{k}\cdot\boldsymbol{\sigma},\label{eq:vr-11}
\end{equation}
where $\hat{\boldsymbol{\beta}}_{k}=\left(\hat{\beta}_{k,1},\hat{\beta}_{k,2},\hat{\beta}_{k,3}\right)$
denotes the control parameters which are essentially the estimates
of $\hat{\boldsymbol{\beta}}$ from the $k$-th iteration. Suppose
$\hat{\boldsymbol{\beta}}_{k}$ deviates from the true value of $\boldsymbol{\beta}$
by $\deltabk[k]$, i.e.,
\begin{equation}
\hat{\boldsymbol{\beta}}_{k}=\boldsymbol{\beta}_{0}+\deltabk[k],\label{eq:vr-12}
\end{equation}
with $\boldsymbol{\beta}_{0}=\left(\beta_{0,1},\beta_{0,2},\beta_{0,3}\right)$
being the true value of $\boldsymbol{\beta}$ and $\deltabk[k]=\left(\delta\beta_{k,1},\delta\beta_{k,2},\delta\beta_{k,3}\right)$
being the estimation errors from the $k$-th iteration, we obtain
the $\deltaes$ of the $\left(k+1\right)$-th iteration as
\begin{equation}
\deltaesk[k+1]=4\left\Vert \deltabk[k]\right\Vert ^{2}.\label{eq:vr-13}
\end{equation}
When the number of trials in the $k$-th iteration, $\tnk[k]$, is
sufficiently large, the central limit theorem guarantees that $\deltabk[k]$
follows a three-dimensional normal distribution asymptotically,
\begin{equation}
\deltabk[k]\sim\mathcal{N}\left(\boldsymbol{0},\covkb[k]\right),\label{eq:vr-14}
\end{equation}
where $\covkb[k]=\left(n_{k}F_{k}^{\left(\boldsymbol{\beta}\right)}\right)^{-1}$.
As $\deltaesk[k+1]$ depends on $\deltabk[k]$, it is also a random
variable. Therefore, we reformulate the optimization objective as
minimizing the expectation value $\deltaeskave[k+1]$,
\begin{equation}
\deltaeskave[k+1]=4\left\langle \delta\beta_{k,1}^{2}+\delta\beta_{k,2}^{2}+\delta\beta_{k,3}^{2}\right\rangle =4\mathrm{Tr}\covkb[k].\label{eq:vr-15}
\end{equation}

\textbf{Optimal evolution time for each trial in one iteration.} Since
the optimal control Hamiltonian requires the knowledge of the unknown
parameters to be estimated, we take an adaptive approach with feedback
to progressively update the control parameters. As time is an important
resource in quantum metrology, we consider a given total evolution
time for the $k$-th iteration, e.g., $T_{k}=n_{k}t_{k}$, where $n_{k}$
and $t_{k}$ are the number of trials and the evolution time per trial
in $k$-th iteration, respectively, and study how to determine the
evolution time $t_{k}$ that minimizes $\deltaeskave[k+1]$.

The control Hamiltonian in the $k$-th iteration depends on the estimated
values of the unknown parameters from the $\left(k-1\right)$-th iteration.
The covariance matrix for the parameters $\beta_{1}$, $\beta_{2}$,
and $\beta_{3}$ is provided in Supplementary Note 7. Applying the
covariance matrix in Eq.~\eqref{eq:vr-15}, we obtain
\begin{equation}
\deltaeskave[k+1]=\frac{1}{T_{k}}\left(\frac{1}{t_{k}}+2t_{k}\left\Vert \deltabk[k-1]\right\Vert ^{2}\csc^{2}\left(\left\Vert \deltabk[k-1]\right\Vert t_{k}\right)\right).\label{eq:oet-1}
\end{equation}

To find the optimal $t_{k}$ that minimizes $\deltaeskave[k+1]$,
we take the derivative of $\deltaeskave[k+1]$ with respect to $t_{k}$.
Let $\prodd=\left\Vert \deltabk[k-1]\right\Vert t_{k}$, a numerical
computation yields the optimal value of $\prodd$ that minimizes $\deltaeskave[k+1]$
as
\begin{equation}
\prodz\approx1.2986.\label{eq:oet-2}
\end{equation}
By using $\left\Vert \deltabk[k-1]\right\Vert ^{2}=\deltaesk[k]/4$
and replacing $\deltaesk[k]$ with $\deltaeskave[k]$, we obtain the
optimal evolution time for each trial in the $k$-th iteration as
\begin{equation}
\toptk[][k]=\frac{2\prodz}{\sqrt{\deltaeskave[k]}}\label{eq:oet-3}
\end{equation}
and a recursive relation for $\deltaeskave[k]$ between two consecutive
iterations,
\begin{equation}
\deltaeskave[k+1]=\frac{\gain[{\prodz}]}{n_{k}}\deltaeskave[k],\label{eq:oet-4}
\end{equation}
where $\gain[x]=1/\left(4x^{2}\right)+\csc^{2}\left(x\right)/2$ and
$\deltaeskave[1]=4\left\Vert \boldsymbol{\beta}_{0}\right\Vert ^{2}$.

\textbf{The scheme with an equal number of trials in each iteration.
}To determine the performance of the adaptive control strategy with
the optimal evolution time derived above, we propose a scheme where
all iterations consist of $\tn$ trials with the respective optimal
evolution time in each iteration. We compare its estimation error
with that of the optimal control strategy which uses the true values
of the unknown parameters.

We define $\vsum[k]=\deltaeskave[k+1]/4$ to represent the sum of
the variances of $\hat{\beta}_{1}$, $\hat{\beta}_{2}$, and $\hat{\beta}_{3}$
in the $k$-th iteration. For the scheme with an equal number $n$
of trials in each iteration, the estimation error after $m$ iterations
is
\begin{equation}
\vsum[m]=\left(\frac{\gain[{\prodz}]}{n}\right)^{m}\vsum[0]\label{eq:tsw-1}
\end{equation}
according to Eq.~\eqref{eq:oet-4}. If the target of estimation error
is $\vsum$, the required number of iterations is given by
\begin{equation}
m=\left\lceil \log_{\frac{\gain[{\prodz}]}{\tn}}\frac{\vsum}{\vsum[0]}\right\rceil ,\label{eq:tsw-2}
\end{equation}
This result shows that the growth of $m$ with decreasing $\vsum$
is slow as it is a logarithm of $V$, implying that the target precision
can be achieved with only a few iterations.

The optimal evolution time for the $k$-th iteration is
\begin{equation}
\toptk[][k]=\prodz\vsum[0]^{-1/2}\left(n/\gain[{\prodz}]\right)^{\left(k-1\right)/2},\label{eq:tsw-3}
\end{equation}
according to Eq.~\eqref{eq:oet-3} and define $\toptk[\mathrm{tot}][m]=\sum_{k=1}^{m}\toptk[][k]$
which is the total evolution time when all the trials are carried
out in parallel for each iteration. To compare with the Heisenberg
limit of the optimal control strategy in Supplementary Note 3, we
derive the relation between $\vsum[m]$ and $\toptk[\mathrm{tot}][m]$,
\begin{equation}
\vsum[m]=\left(\frac{\gain[{\prodz}]}{\tn}\right)^{m}\left(\prodz\frac{1-\sqrt{\frac{\tn}{\gain[{\prodz}]}}^{m}}{1-\sqrt{\frac{\tn}{\gain[{\prodz}]}}}\right)^{2}\toptk[\mathrm{tot}][m]^{-2}.\label{eq:tsw-4}
\end{equation}
For $\sqrt{\tn/\gain[{\prodz}]}\gg1$, Eq.~\eqref{eq:tsw-4} simplifies
to
\begin{equation}
\vsum[m]\approx\frac{\prodz^{2}\gain[{\prodz}]}{n\toptk[\mathrm{tot}][m]^{2}}.\label{eq:tsw-5}
\end{equation}
This result confirms that the Heisenberg limit can be achieved by
the above adaptive control approach with the evolution time optimized
in each iteration.

For the optimal control Hamiltonian, $\hcont=-\boldsymbol{\beta}_{0}\cdot\boldsymbol{\sigma}$,
Supplementary Note 3 shows that the covariance matrix for the parameters
$\beta_{1}$, $\beta_{2}$, and $\beta_{3}$ is
\begin{equation}
\covkb[\mathrm{oc}]=\frac{1}{4\tnoc\tkoc^{2}}I_{3\times3},\label{eq:tsw-6}
\end{equation}
where $I_{3\times3}$ is the three-dimensional identity matrix, and
the total variance of the estimators $\hat{\beta}_{1}$, $\hat{\beta}_{2}$,
and $\hat{\beta}_{3}$ is therefore
\begin{equation}
\vsum[\mathrm{oc}]=\frac{3}{4\tnoc\tkoc^{2}}.\label{eq:tsw-7}
\end{equation}
Compared to this optimal control strategy with precise control parameters,
$\vsum[m]$ is only $4\prodz^{2}\gain[{\prodz}]/3\approx1.55$ times
larger, implying the estimation precision of this adaptive control
strategy achieves the optimum up to an overall factor.

The above result is obtained for the parameters in the Pauli basis,
$\beta_{1}$, $\beta_{2}$, and $\beta_{3}$, rather than the original
parameters $\alpha_{i}$ in the Hamiltonian. The Supplementary Note
8 derives the relation between estimation variances of the original
parameter with adaptive and optimal control, from which we obtain
\begin{equation}
\left\langle \delta^{2}\widehat{\alpha}_{i}\right\rangle _{m}\leq\left(4\prodzs\csc^{2}\left(\prodz\right)-1\right)\left\langle \delta^{2}\widehat{\alpha}_{i}\right\rangle _{\mathrm{oc}}\approx6.27\left\langle \delta^{2}\widehat{\alpha}_{i}\right\rangle _{\mathrm{oc}},\label{eq:tsw-8}
\end{equation}
where $\left\langle \delta^{2}\widehat{\alpha}_{i}\right\rangle _{m}$
and $\left\langle \delta^{2}\widehat{\alpha}_{i}\right\rangle _{\mathrm{oc}}$
represent the estimation variances of $\widehat{\alpha}_{i}$ with
adaptive and optimal control, respectively. This indicates that our
adaptive control strategy can work for arbitrary parameters of a Hamiltonian
in general and the estimation precision for any unknown parameter
in the Hamiltonian can also achieve the optimum up to a factor of
constant order.

\textbf{Discussion.} The optimal evolution time $\toptk[][k]$ in
the $k$-th iteration is derived based on the expectation value of
$\deltaesk[k]$. In practical experiments, the measurement results
of the $\left(k-1\right)$-th iteration can be random, so $\deltaesk[k]$,
which is obtained from the $\left(k-1\right)$-th iteration, can also
be random accordingly. Hence, the real value of $\deltaesk[k]$ can
deviate from its expectation value in practice, which may affect the
estimation precision at the $k$-th iteration. In the Methods, we
study the effect of this randomness on the optimal evolution time
scheme and the robustness of this scheme. In particular, we show it
is more probable that the estimation precision can benefit from such
deviation in $\deltaesk[k]$ and perform better than the case with
the expectation value of $\deltaesk[k]$. Therefore, the random deviation
in $\deltaesk[k]$ can actually be favorable to this adaptive feedback
control strategy.

\section*{Methods}

\begin{figure}
\includegraphics[scale=0.4]{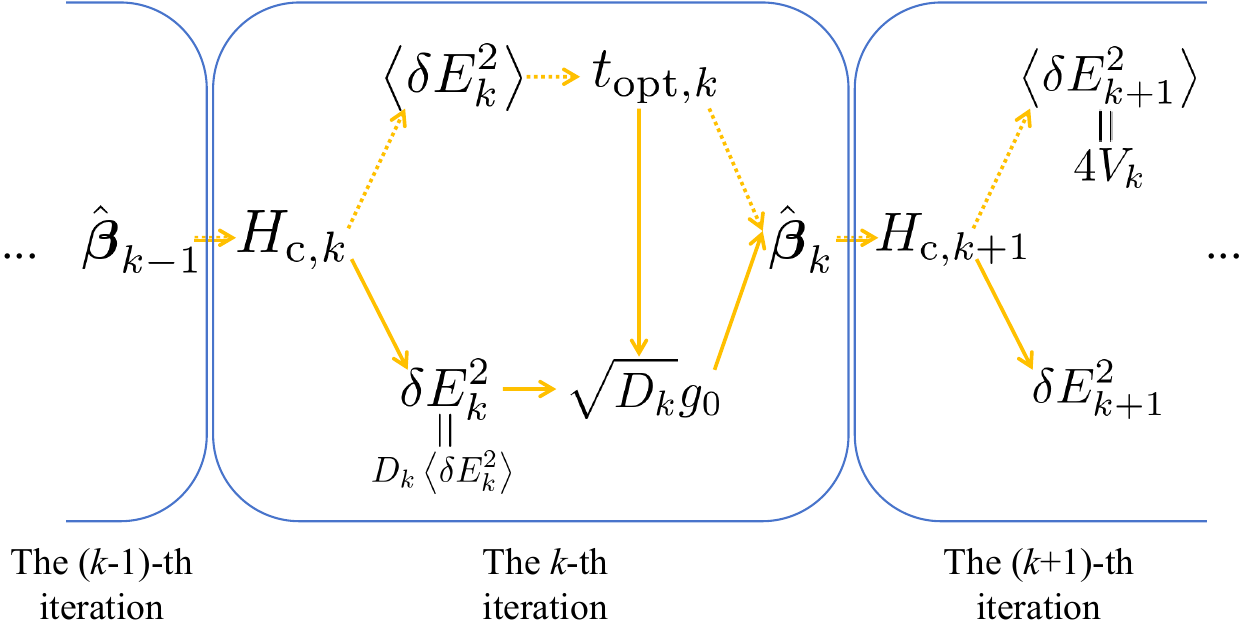}

\caption{\textbf{Relations between different physical quantities.} Arrows
schematically denote the relations between different physical quantities
occurred in the proposed optimal evolution time scheme, pointing from
one quantity to the derived quantity. The upper section, linked by
dashed arrows, depicts the no-deviation cases with the errors of all
control parameters averaged. The lower section, linked by solid arrows,
depicts the practical cases where the errors of the control parameters
have random deviation from their average values, manifested by the
deviation factor $D_{k}$ for the $k$-th iteration.}\label{fig:physical_quantities}
\end{figure}

\begin{figure}
\subfloat[\label{fig:robustness_analysis-a}]{\includegraphics[scale=0.35]{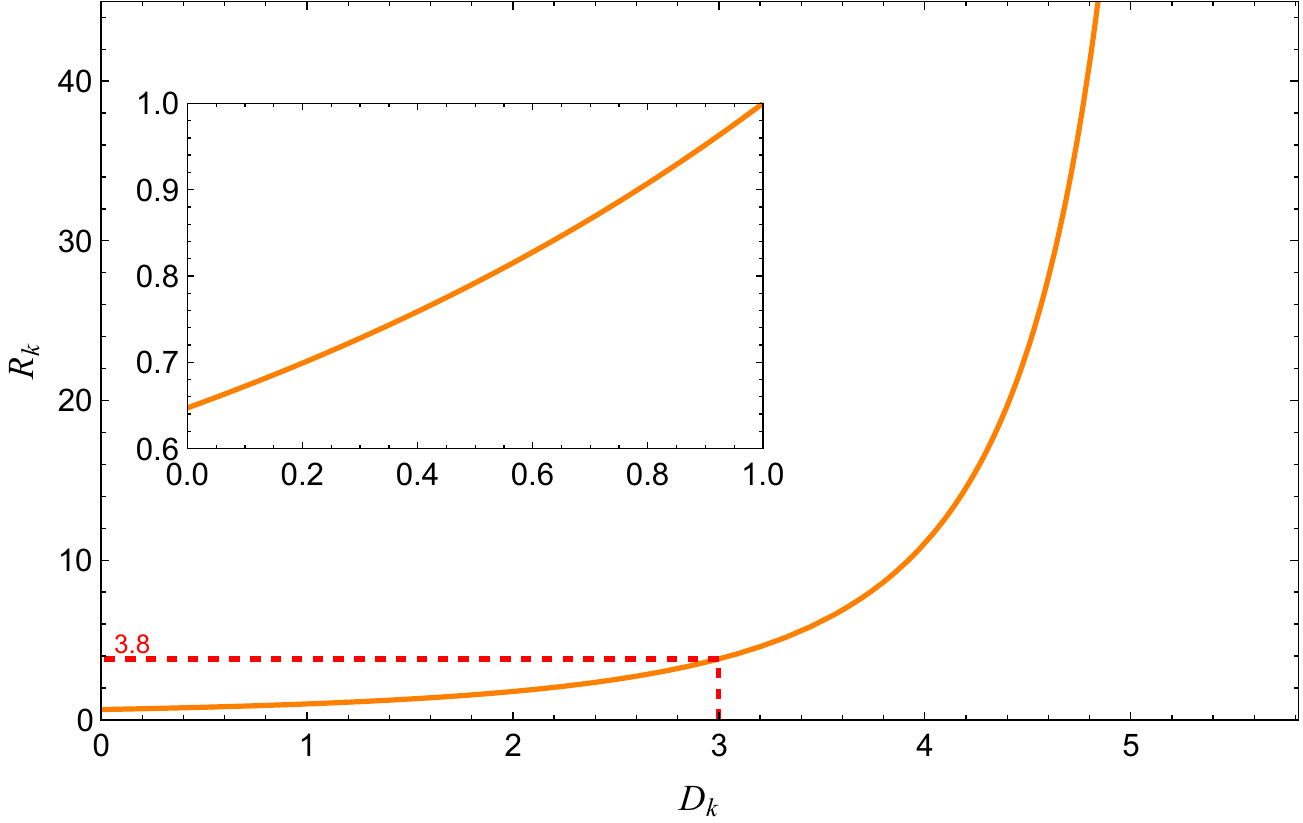}

}

\subfloat[\label{fig:robustness_analysis-b}]{\includegraphics[scale=0.35]{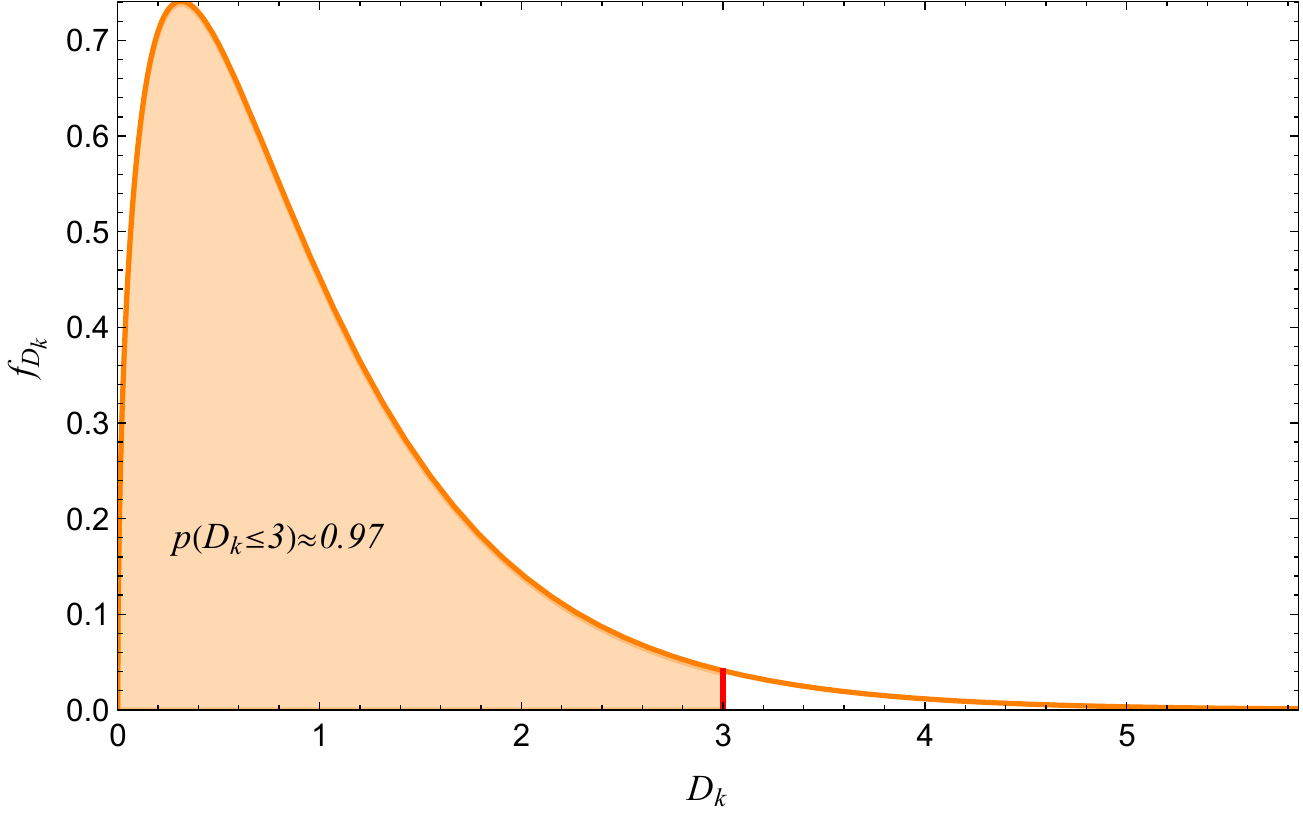}

}

\caption{\textbf{Robustness of a single iteration against deviation in the
errors of control parameters.} Fig. (a) illustrates the effect of
the deviation factor on the estimation precision. The estimation precision
decreases as $\protect\dev$ increases. When $\protect\dev$ is sufficiently
low, the real estimation precision approaches the optimal precision
with precise control parameters, so the ratio $R_{k}$ between the
real estimation precision to the estimation precision with average
errors in the control parameters drops below $1$, as shown by the
left panel in this figure. In the region indicated by the red dashed
line in this figure, the estimation precision is almost insensitive
to the random deviation of the errors of control parameters. Fig.
(b) shows the probability density function of $\protect\dev$, which
suggests $\protect\dev$ lies in an interval where the estimation
precision is close to that with average errors in the control parameters
with a high probability, indicating strong robustness of the optimal
evolution time scheme against the deviation in the errors of control
parameters.}\label{fig:robustness_analysis}
\end{figure}

\textbf{Robustness Analysis.} To facilitate the following analysis,
Fig. \ref{fig:physical_quantities} schematically depicts the relations
between the physical quantities in the optimal evolution time scheme,
using dashed arrows for the no-deviation case with the errors of the
control parameters averaged and the solid arrows for the practical
cases with random deviations in the average errors of the control
parameters. To explicitly characterize the deviation of $\deltaesk[k]$
from its expectation value, we introduce a deviation factor $\dev$,
which is also random due to the randomness of $\deltaesk[k]$,
\begin{equation}
\deltaesk[k]=\dev\deltaeskave[k].\label{eq:ra-1}
\end{equation}
We use the evolution time $\toptk[][k]$, which is derived based on
the mean of $\deltaesk[k]$ according to Eq.~\eqref{eq:oet-3}, to
perform the $k$-th iteration. If $\deltaesk[k]$ deviates from its
mean in the experiment, Eq.~\eqref{eq:oet-3} shows that the deviation
factor $\dev$ scales $\prodz$ to $\sqrt{\dev}\prodz$, and thus
the recursive relation Eq.~\eqref{eq:oet-4} becomes
\begin{equation}
\deltaeskave[k+1]_{\dev}=\frac{\gain[{\sqrt{\dev}\prodz}]}{n_{k}}\dev\deltaeskave[k].\label{eq:ra-2}
\end{equation}
The impact of the deviation factor $\dev$ on the estimation precision
at the $k$-th iteration is characterized by the ratio
\begin{equation}
R_{k}=\frac{\vsum[{k,\dev}]}{\vsum[k]}=\frac{\dev G\left(\sqrt{\dev}\prodz\right)}{G\left(\prodz\right)}\label{eq:ra-3}
\end{equation}
as illustrated in Fig. \ref{fig:robustness_analysis-a}. The estimation
precision decreases as the deviation increases for $0<\dev<\left(\pi/\prodz\right)^{2}$.
When $0<\dev<1$, where the control Hamiltonian obtained from the
estimate in the $\left(k-1\right)$-th iteration leads to a $\deltaesk[k]$
smaller than its mean, implying that the estimate from the $\left(k-1\right)$-th
iteration is better than the average, we have $R_{k}<1$. As $\dev\rightarrow0$,
$\vsum[{k,\dev}]\rightarrow3/\left(4n_{k}\toptk[][k]^{2}\right)$,
which is exactly the bound given in Supplementary Note 3. For $1\leq\dev<\left(\pi/\prodz\right)^{2}$,
indicating that the estimate from the $\left(k-1\right)$-th iteration
is worse than the average, it follows that $R_{k}\geq1$.

The deviation factor $\dev$ follows a generalized $\chi^{2}$ distribution
\begin{equation}
\dev=\frac{\chi_{1}^{2}\left(1\right)+\prodzs\csc^{2}\left(\prodz\right)\left(\chi_{2}^{2}\left(1\right)+\chi_{3}^{2}\left(1\right)\right)}{2\prodzs\csc^{2}\left(\prodz\right)+1},\label{eq:ra-4}
\end{equation}
where $\chi_{1}^{2}\left(1\right)$, $\chi_{2}^{2}\left(1\right)$,
and $\chi_{3}^{2}\left(1\right)$ are squares of independent standard
normal random variables. Detail of the derivation can be found in
Supplementary Note 9. The probability density $f_{\dev}$ of the deviation
factor $\dev$ is shown in Fig. \ref{fig:robustness_analysis-b}.
Fig. \ref{fig:robustness_analysis} suggests that the deviation factor
$\dev$ lies most likely in a region where $R_{k}$ is almost insensitive
to $\dev$ and close to $1$, demonstrating strong robustness of the
estimation precision against the deviation in the errors of control
parameters in a single iteration.

In practice, the deviation factor $\dev$ modifies the optimal evolution
time $\toptk[][k]$, which is determined based on the mean of $\deltaesk[k]$,
to $\toptk[][k]/\sqrt{\dev}$. Therefore, an evolution time modified
procedure is required. Suppose the estimate obtained from the $\left(k-1\right)$-th
iteration is $\boldsymbol{\beta}_{k-1}^{'}$, which determines the
control Hamiltonian at the $k$-th iteration. By continuing to repeat
the trials in the $\left(k-1\right)$-th iteration, a more precise
estimate $\boldsymbol{\beta}_{0}^{'}$ can be obtained. The modified
evolution time $\widetilde{\toptk[][k]}$ is determined by $\left\Vert \boldsymbol{\beta}_{k-1}^{'}-\boldsymbol{\beta}_{0}^{'}\right\Vert \widetilde{\toptk[][k]}=\Pi_{0}$,
after which the $k$-th iteration proceeds with $\hcontkth[k]=-\boldsymbol{\beta}_{k-1}^{'}\cdot\boldsymbol{\sigma}$
and $\widetilde{\toptk[][k]}$.

We now consider the effect of the deviation factors on the estimation
precision of the estimation process consisting of $m$ iterations
with the evolution time modified procedure. Since the evolution time
of each iteration is modified to the optimal evolution time, $\prodz$
is not scaled by the deviation factor $\dev$. To ensure a effective
comparison with the same total evolution time, we adopt the equivalent
form of Eq.~\eqref{eq:oet-4},
\begin{equation}
\deltaeskave[k+1]=\frac{2\prodz\gain[{\prodz}]}{T_{k}}\sqrt{\deltaeskave[k]},\label{eq:ra-5}
\end{equation}
which yields
\begin{equation}
\widetilde{\deltaeskave[k+1]}_{\dev}=\frac{2\prodz\gain[{\prodz}]}{T_{k}}\sqrt{\dev\widetilde{\deltaeskave[k]}_{\dev[k-1]}},\label{eq:ra-6}
\end{equation}
where $\dev[1]=1$ and $\widetilde{\deltaeskave[1]}_{\dev[0]}=\deltaeskave[1]$.
The effect of the deviation factors on the estimation precision at
the entire estimation process is characterized by the ratio
\begin{equation}
\begin{array}{rl}
\widetilde{R}_{\mathrm{tot},m}= & \frac{\vsumt[{m,\dev}]}{\vsum[m]}\\
= & \prod_{k=1}^{m}\dev^{\frac{1}{2^{m-\left(k-1\right)}}},
\end{array}\label{eq:ra-7}
\end{equation}
where $\vsumt[{m,\dev}]=\widetilde{\deltaeskave[m+1]}_{\dev}/4$.

\begin{figure}
\includegraphics[scale=0.33]{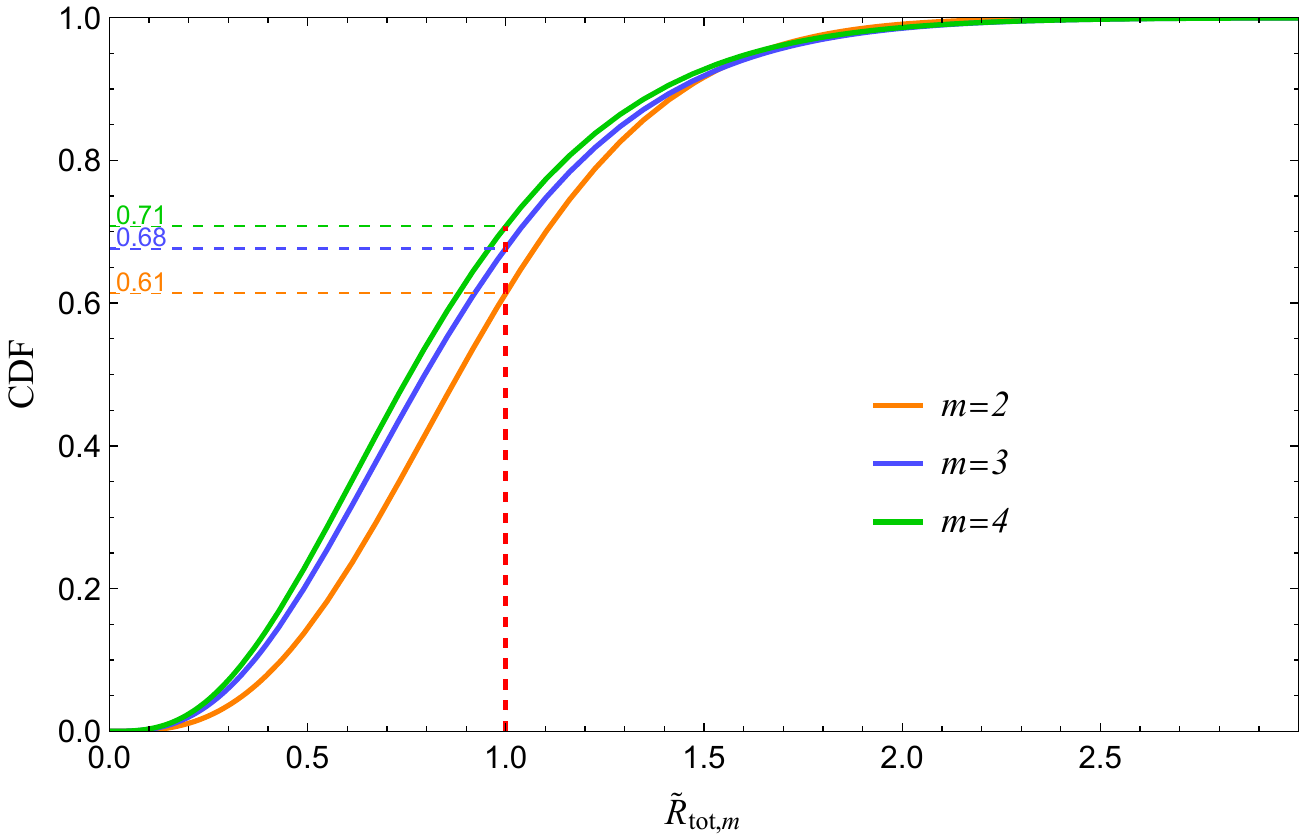}

\caption{\textbf{Robustness of the optimal evolution time scheme.} These figures
illustrate the effect of the deviation in $\protect\deltaesk[k]$
on the estimation precision for a total of two (orange curve), three
(blue curve), and four (green curve) iterations. The orange, blue,
and green dashed lines plot the probabilities that the precision with
the deviation in $\protect\deltaesk[k]$ surpasses that without deviation
for different iteration numbers. They all exceed $50\%$ and increase
with the number of iterations, implying that the real estimation precision
actually benefits from the deviation in $\protect\deltaesk[k]$ and
becomes better than the expected estimation precision.}\label{fig:robustness_analysis2}
\end{figure}

Fig. \ref{fig:robustness_analysis2} shows the cumulative distribution
functions of $\widetilde{R}_{\mathrm{tot},m}$ for $m=2,3$, and $4$.
A surprising result from the figure is that the probabilities that
the estimation variance with deviation in $\deltaesk[k]$ surpasses
that without deviation exceed $50\%$ and increase with the number
of iterations, suggesting that the real estimation precision is more
likely to benefit from the deviation in $\deltaesk[k]$ and becomes
better than the expected estimation precision.

\section*{Data availability}

The code and data used in this work are available upon request to
the corresponding author.

\section*{Acknowledgments}

This work is supported by National Natural Science Foundation of China
(Grant No. 12075323), the Natural Science Foundation of Guangdong
Province of China (Grant No. 2025A1515011440) and the Innovation Program
for Quantum Science and Technology (Grant No. 2021ZD0300702).

\section*{Author contributions}

Q.W. initiated this work, and carried out the main calculations. S.P.
participated in scientific discussions, and assisted with the calculations.
Both authors contributed to the writing of the manuscript.

\section*{Competing financial interests}

The authors declare no competing financial interests.

\bibliographystyle{apsrev4-2}
\bibliography{reference}

\newpage\setcounter{equation}{0} \newcounter{suppnote} 
\global\long\def\theequation{S\arabic{equation}}%
 \onecolumngrid \setcounter{enumiv}{0}

\refstepcounter{suppnote}

\section*{Supplementary Note 1. Eliminating measurement trade-off through system
extension}\label{sec:note1}

This Supplementary Note proves that the system extension scheme with
the initial state being a maximally entangled state eliminates the
measurement tradeoff.

Suppose a $d$-dimensional system governed by a Hamiltonian $\ha$,
we introduce an ancillary system with dimension no less than $d$,
whose Hamiltonian is the identity operator $\iA$. The initial state
is prepared as an arbitrary pure state of the joint system, denoted
as $\psiz=\sum_{l=0}^{d-1}\lambdai[l]\left|\bv[\mathrm{P}]\right\rangle \otimes\left|\bv[\mathrm{A}]\right\rangle $,
where $\left\{ \left|\bv[\mathrm{P}]\right\rangle \left|\ l=0,1,\ldots,d-1\right.\right\} $
forms a complete orthonormal basis of the system, $\left\{ \left|\bv[\mathrm{A}]\right\rangle \left|\ l=0,1,\ldots,d-1\right.\right\} $
is a set of mutually orthogonal basis vectors for the ancillary system,
and the non-negative coefficients $\lambdai[l]$ satisfy $\sum_{l=0}^{d-1}\lambdai[l]^{2}=1$.
The Hamiltonian of the joint system is $\ha\otimes\iA$, so the generator
of the infinitesimal translation of $\unita\otimes\iA$ with respect
to the parameter $\alpha_{i}$ is given by $\gen[i]\otimes\iA$, where
$\gen[i]=-i\left(\pd[i]\unita^{\dagger}\right)\unita$ and $\unita=\exp\left(-i\ha t\right)$.
From Eq.~(7) of the main manuscript, we obtain
\begin{equation}
\psizct\gen[i]\gen[j]\otimes\iA\psiz=\psizct\gen[j]\gen[i]\otimes\iA\psiz.\label{eq:emt-1}
\end{equation}
Calculating both sides of the equation separately, we obtain
\begin{equation}
\psizct\gen[i]\gen[j]\otimes\iA\psiz=\sum_{l=0}^{d-1}\lambdai[l]^{2}\left\langle \bv[\mathrm{P}]\right|\gen[i]\gen[j]\left|\bv[\mathrm{P}]\right\rangle ,\label{eq:emt-2}
\end{equation}
and
\begin{equation}
\psizct\gen[j]\gen[i]\otimes\iA\psiz=\sum_{l=0}^{d-1}\lambdai[l]^{2}\left\langle \bv[\mathrm{P}]\right|\gen[j]\gen[i]\left|\bv[\mathrm{P}]\right\rangle .\label{eq:emt-3}
\end{equation}
Since $\gen[i]$ and $\gen[j]$ generally do not commute, Eq.~\eqref{eq:emt-1}
holds only when $\lambdai[l]=\frac{1}{\sqrt{d}}$ due to the cyclic
property of the trace operator. If the dimension of the ancillary
system is equal to that of the system, we conclude that the weak commutativity
condition is satisfied when the initial state is the maximally entangled
state.

\refstepcounter{suppnote}

\section*{Supplementary Note 2. Eliminating initial-state trade-off in two-dimensional
systems}\label{sec:note2}

This supplementary note proves that, for a two-dimensional system
with system extension, choosing the maximally entangled state as the
initial state makes the quantum Fisher information matrix optimal.

Suppose the joint Hamiltonian of a two-dimensional system and a two-dimensional
ancillary system is $\ha\otimes\iA$ and the initial state is $\psiz=\sqrt{x}\left|\bvdtf[\mathrm{P}]\bvdtf[\mathrm{A}]\right\rangle +\sqrt{1-x}\left|\bvdts[\mathrm{P}]\bvdts[\mathrm{A}]\right\rangle $
($0\leq x\leq1$), where $\left\{ \left|\bvdtf[\mathrm{P}]\right\rangle ,\left|\bvdts[\mathrm{P}]\right\rangle \right\} $
and $\left\{ \left|\bvdtf[\mathrm{A}]\right\rangle ,\left|\bvdts[\mathrm{A}]\right\rangle \right\} $
are sets of complete orthonormal basis for the system and ancillary
system, respectively. According to Eq.~(4) in the main manuscript,
the entries of quantum Fisher information matrix of the finial state
are given by
\begin{equation}
\begin{array}{rl}
F_{ij}= & 4\left\{ \mathrm{Re}\left[x\left\langle \bvdtf[\mathrm{P}]\right|\gen[i]\gen[j]\left|\bvdtf[\mathrm{P}]\right\rangle +\left(1-x\right)\left\langle \bvdts[\mathrm{P}]\right|\gen[i]\gen[j]\left|\bvdts[\mathrm{P}]\right\rangle \right]\right.\\
 & \left.-\left[x\left\langle \bvdtf[\mathrm{P}]\right|\gen[i]\left|\bvdtf[\mathrm{P}]\right\rangle +\left(1-x\right)\left\langle \bvdts[\mathrm{P}]\right|\gen[i]\left|\bvdts[\mathrm{P}]\right\rangle \right]\left[x\left\langle \bvdtf[\mathrm{P}]\right|\gen[j]\left|\bvdtf[\mathrm{P}]\right\rangle +\left(1-x\right)\left\langle \bvdts[\mathrm{P}]\right|\gen[j]\left|\bvdts[\mathrm{P}]\right\rangle \right]\right\} .
\end{array}\label{eq:eit-1}
\end{equation}
Denote the matrix representation of $\gen[i]$ in the basis $\left\{ \left|\bvdtf[\mathrm{P}]\right\rangle ,\left|\bvdts[\mathrm{P}]\right\rangle \right\} $
as
\begin{equation}
\genm[i]=\left\{ \begin{array}{rl}
\genm[i,11] & \genm[i,12]\\
\genm[i,21] & \genm[i,22]
\end{array}\right\} ,\label{eq:eit-2}
\end{equation}
we have
\begin{equation}
\begin{array}{rl}
F_{ij}= & 4\left\{ \mathrm{Re}\left[x\left(\genm[i,11]\genm[j,11]+\genm[i,12]\genm[j,21]\right)+\left(1-x\right)\left(\genm[i,21]\genm[j,12]+\genm[i,22]\genm[j,22]\right)\right]\right.\\
 & \left.-\left[x\genm[i,11]+\left(1-x\right)\genm[i,22]\right]\left[x\genm[j,11]+\left(1-x\right)\genm[j,22]\right]\right\} .
\end{array}\label{eq:eit-3}
\end{equation}

We define $\deltaf=\left.F\right|_{x=1/2}-F$, the elements of $\deltaf$
are given by
\begin{equation}
\deltaf_{ij}=4\left(\genm[i,11]-\genm[i,22]\right)\left(\genm[j,11]-\genm[j,22]\right)\left(x-\frac{1}{2}\right)^{2}.\label{eq:eit-4}
\end{equation}
For both the two-parameter and three-parameter cases, all principal
minors of $\deltaf$ are non-negative, which indicates that $\deltaf$
is positive semi-definite. Therefore, the maximally entangled state
is the optimal pure state.

Let $\rho_{0}=\underset{i}{\sum}p_{i}\left|\varphi_{i}\right\rangle \left\langle \varphi_{i}\right|$
be an arbitrary mixed initial state. Under the unitary evolution $\unita\otimes\iA$,
the final state is $\rho_{t}=\underset{i}{\sum}p_{i}\left(\unita\otimes\iA\right)\left|\varphi_{i}\right\rangle \left\langle \varphi_{i}\right|\left(U_{\boldsymbol{\alpha}}^{\dagger}\otimes\iA\right)$.
According to the convexity of the quantum Fisher information matrix
\citep{liu2019quantum}, we obtain
\begin{equation}
\begin{array}{rl}
F\left(\rho_{t}\right)\leq & \underset{i}{\sum}p_{i}F\left[\left(\unita\otimes\iA\right)\left|\varphi_{i}\right\rangle \left\langle \varphi_{i}\right|\left(U_{\boldsymbol{\alpha}}^{\dagger}\otimes\iA\right)\right]\\
\leq & \left.\underset{i}{\sum}p_{i}F\left[\left(\unita\otimes\iA\right)\psiz\psizct\left(U_{\boldsymbol{\alpha}}^{\dagger}\otimes\iA\right)\right]\right|_{x=\frac{1}{2}}\\
= & \left.F\right|_{x=\frac{1}{2}},
\end{array}\label{eq:eit-5}
\end{equation}
demonstrating that the maximally entangled initial state is optimal.

\refstepcounter{suppnote}

\section*{Supplementary Note 3. Achieving Heisenberg scaling via Hamiltonian
control}\label{sec:note3}

This supplementary note proves that using the reverse of the initial
Hamiltonian as the control Hamiltonian enables the parameter estimation
precision to reach the Heisenberg limit.

Consider a two-dimensional system with Hamiltonian $\ha$ and a two-dimensional
ancillary system with Hamiltonian $\iA$. The probe and ancilla are
initialized in a maximally entangled state and evolve under the joint
Hamiltonian $\ha\otimes\iA$. The entries of quantum Fisher information
matrix for the final evolved state are given by
\begin{equation}
F_{ij}=2\mathrm{Tr}\left(\gent[i]\gent[j]\right)-t^{2}\mathrm{Tr}\left(\pd[i]\ha\right)\mathrm{Tr}\left(\pd[j]\ha\right),\label{eq:ahs-1}
\end{equation}
where $\gent[i]$ can be expressed as
\begin{equation}
\gent[i]=\int_{0}^{t}\mathrm{e}^{i\ha\tau}\left(\partial_{i}\ha\right)\mathrm{e}^{-i\ha\tau}d\tau\label{eq:ahs-2}
\end{equation}
and we have used
\[
\mathrm{Tr}\left(\gent[i]\right)=t\mathrm{Tr}\left(\pd[i]\ha\right).
\]
Suppose the initial Hamiltonian of the system is $\hainit$, then
the control Hamiltonian is $H_{\mathrm{c}}=-\hainit$. Here, $\hainit$
is parameter-dependent, while $H_{\mathrm{c}}$ is parameter-independent.
The total Hamiltonian $\ha=\hainit+H_{\mathrm{c}}$ becomes a zero
operator, yielding $\gent[i]=t\pd[i]\hainit$ , and hence
\begin{equation}
F_{ij}^{\left(\mathrm{c}\right)}=t^{2}\left\{ 2\mathrm{Tr}\left[\left(\pd[i]\hainit\right)\left(\pd[j]\hainit\right)\right]-\mathrm{Tr}\left(\pd[i]\hainit\right)\mathrm{Tr}\left(\pd[j]\hainit\right)\right\} .\label{eq:ahs-3}
\end{equation}
The estimator variance of the parameter $\alpha_{i}$ is given by
\begin{equation}
\left\langle \delta^{2}\widehat{\alpha}_{i}\right\rangle =\left[\left(nF^{\left(\mathrm{c}\right)}\right)^{-1}\right]_{ii}\propto\frac{1}{nt^{2}},\label{eq:ahs-4}
\end{equation}
where $\left\langle \delta^{2}\widehat{\alpha}_{i}\right\rangle \coloneqq\mathrm{E}\left[\left(\widehat{\alpha}_{i}/\pd[\alpha_{i}]\mathrm{E}\left(\widehat{\alpha}_{i}\right)-\alpha_{i}\right)^{2}\right]$.

\refstepcounter{suppnote}

\section*{Supplementary Note 4. Time dependence of quantum Fisher information
matrix}\label{sec:note4}

In this Supplementary Note, we derive the explicit time dependence
of quantum Fisher information matrix.

To obtain the explicit time dependence of the quantum Fisher information
matrix, we first solve for the explicit time dependence of $\gent[i]$
in Eq.~\eqref{eq:ahs-2}. Let $Y\left(\tau\right)=\mathrm{e}^{iH_{\boldsymbol{\alpha}}\tau}\left(\partial_{i}H_{\boldsymbol{\alpha}}\right)\mathrm{e}^{-iH_{\boldsymbol{\alpha}}\tau}$,
we obtain
\begin{equation}
\frac{\partial Y\left(\tau\right)}{\partial\tau}=i\left[H_{\boldsymbol{\alpha}},Y\right],\label{eq:tdo-1}
\end{equation}
where $Y\left(0\right)=\partial_{i}H_{\boldsymbol{\alpha}}$. The
operator $\mathcal{H}\left(\cdot\right)\coloneqq\left[H_{\boldsymbol{\alpha}},\cdot\right]$
is an Hermitian superoperator with four eigenvalues $\varLambda_{1},\varLambda_{2},\varLambda_{3},\varLambda_{4}$,
where $\varLambda_{k}=0$ for $k=1,...,r$ and $\varLambda_{k}\neq0$
for $k=r+1,...,4$. The corresponding eigenvectors are $\varGamma_{1},\varGamma_{2},\varGamma_{3},\varGamma_{4}$,
which satisfy $\mathrm{Tr}\left[\varGamma_{i}^{\dagger}\varGamma_{j}\right]=\delta_{ij}$.
We can decompose $Y\left(0\right)$ based on $\left\{ \varGamma_{k}\right\} $
as
\begin{equation}
Y\left(0\right)=\sum_{k=1}^{4}c_{k}\varGamma_{k},\label{eq:tdo-2}
\end{equation}
where $c_{k}=\mathrm{Tr}\left(\varGamma_{k}^{\dagger}\partial_{i}H_{\boldsymbol{\alpha}}\right)$.
The solution for $Y\left(\tau\right)$ is
\begin{equation}
Y\left(\tau\right)=\sum_{k=1}^{4}\mathrm{Tr}\left(\varGamma_{k}^{\dagger}\partial_{i}H_{\boldsymbol{\alpha}}\right)\exp\left(i\varLambda_{k}\tau\right)\varGamma_{k},\label{eq:tdo-3}
\end{equation}
leading to
\begin{equation}
\gent[i]=t\sum_{k=1}^{r}\mathrm{Tr}\left(\varGamma_{k}^{\dagger}\partial_{i}H_{\boldsymbol{\alpha}}\right)\varGamma_{k}-i\sum_{k=r+1}^{4}\frac{\exp\left(i\varLambda_{k}t\right)-1}{\varLambda_{k}}\mathrm{Tr}\left(\varGamma_{k}^{\dagger}\partial_{i}H_{\boldsymbol{\alpha}}\right)\varGamma_{k}.\label{eq:tdo-4}
\end{equation}
Suppose $H_{\boldsymbol{\alpha}}$ has two eigenvalues $\eigen[0]$
and $\eigen[1]$ with $\eigen[0]\neq\eigen[1]$, and two corresponding
eigenvectors $\left|\eigen[0]\right\rangle $ and $\left|\eigen[1]\right\rangle $,
the eigenvalues and eigenvectors of $\mathcal{H}\left(\cdot\right)$
are
\begin{equation}
\begin{array}{ll}
\varLambda_{1}=0 & \varGamma_{1}=\left|\eigen[0]\right\rangle \left\langle \eigen[0]\right|,\\
\varLambda_{2}=0 & \varGamma_{2}=\left|\eigen[1]\right\rangle \left\langle \eigen[1]\right|,\\
\varLambda_{3}=\eigen[0]-\eigen[1] & \varGamma_{3}=\left|\eigen[0]\right\rangle \left\langle \eigen[1]\right|,\\
\varLambda_{4}=\eigen[1]-\eigen[0] & \varGamma_{4}=\left|\eigen[1]\right\rangle \left\langle \eigen[0]\right|.
\end{array}\label{eq:tdo-5}
\end{equation}
The solutions for $\mathrm{Tr}\left(\varGamma_{k}^{\dagger}\partial_{i}H_{\boldsymbol{\alpha}}\right)$
are
\begin{equation}
\begin{array}{rl}
\mathrm{Tr}\left(\varGamma_{1}^{\dagger}\partial_{i}H_{\boldsymbol{\alpha}}\right)= & \partial_{i}E_{0},\\
\mathrm{Tr}\left(\varGamma_{2}^{\dagger}\partial_{i}H_{\boldsymbol{\alpha}}\right)= & \partial_{i}\eigen[1],\\
\mathrm{Tr}\left(\varGamma_{3}^{\dagger}\partial_{i}H_{\boldsymbol{\alpha}}\right)= & \left(\eigen[1]-\eigen[0]\right)\left\langle \eigen[0]\right|\left.\partial_{i}\eigen[1]\right\rangle ,\\
\mathrm{Tr}\left(\varGamma_{4}^{\dagger}\partial_{i}H_{\boldsymbol{\alpha}}\right)= & \left(\eigen[0]-\eigen[1]\right)\left\langle \eigen[1]\right|\left.\partial_{i}\eigen[0]\right\rangle .
\end{array}\label{eq:tdo-6}
\end{equation}
Substituting Eq.~\eqref{eq:tdo-5} and Eq.~\eqref{eq:tdo-6} into
Eq.~\eqref{eq:tdo-4}, we obtain
\begin{equation}
\gent[i]=\sum_{l=0}^{1}t\left(\partial_{i}\eigen[l]\right)\left|\eigen[l]\right\rangle \left\langle \eigen[l]\right|+i\left[\exp\left(i\left(\eigen[l]-\eigen[1-l]\right)t\right)-1\right]\left\langle \eigen[l]\right|\left.\partial_{i}\eigen[1-l]\right\rangle \left|\eigen[l]\right\rangle \left\langle \eigen[1-l]\right|.\label{eq:tdo-7}
\end{equation}
Finally, substituting Eq.~\eqref{eq:tdo-7} into Eq.~\eqref{eq:ahs-1}
yields
\begin{equation}
F_{ij}\left(t\right)=\sum_{l=0}^{1}t^{2}\left[\left(\pd[i]E_{l}\right)\left(\pd[j]E_{l}\right)-\left(\pd[i]E_{l}\right)\left(\pd[j]E_{1-l}\right)\right]-8\sin^{2}\left(\frac{\left(E_{l}-E_{1-l}\right)t}{2}\right)\left\langle E_{l}\right|\left.\pd[i]E_{1-l}\right\rangle \left\langle E_{1-l}\right|\left.\pd[j]E_{l}\right\rangle .\label{eq:tdo-8}
\end{equation}

\refstepcounter{suppnote}

\section*{Supplementary Note 5. Time dependence of estimator variances}\label{sec:note5}

This Supplementary Note provides an explicit expression for the estimator
variances and analyses its dependence on time.

The quantum Fisher information matrix can be directly obtained from
Eq.~\eqref{eq:tdo-8}. The variance of $\widehat{\alpha}_{1}$, $\widehat{\alpha}_{2}$,
and $\widehat{\alpha}_{3}$ are

\begin{equation}
\left\langle \delta^{2}\widehat{\alpha}_{1}\right\rangle =\frac{t^{2}\csc^{2}\left(\frac{1}{2}\deltae t\right)\left[\left(\realvei[2]\pd[3]\deltae-\realvei[3]\pd[2]\deltae\right)^{2}+\left(\imvei[2]\pd[3]\deltae-\imvei[3]\pd[2]\deltae\right)^{2}\right]+16\left(\realvei[3]\imvei[2]-\realvei[2]\imvei[3]\right)^{2}}{16nt^{2}\left[\realvei[1]\left(\imvei[3]\pd[2]\deltae-\imvei[2]\pd[3]\deltae\right)+\realvei[2]\left(\imvei[1]\pd[3]\deltae-\imvei[3]\pd[1]\deltae\right)+\realvei[3]\left(\imvei[2]\pd[1]\deltae-\imvei[1]\pd[2]\deltae\right)\right]^{2}},\label{eq:tdoe-1}
\end{equation}
\begin{equation}
\left\langle \delta^{2}\widehat{\alpha}_{2}\right\rangle =\frac{t^{2}\csc^{2}\left(\frac{1}{2}\deltae t\right)\left[\left(\realvei[1]\pd[3]\deltae-\realvei[3]\pd[1]\deltae\right)^{2}+\left(\imvei[1]\pd[3]\deltae-\imvei[3]\pd[1]\deltae\right)^{2}\right]+16\left(\realvei[3]\imvei[1]-\realvei[1]\imvei[3]\right)^{2}}{16nt^{2}\left[\realvei[1]\left(\imvei[3]\pd[2]\deltae-\imvei[2]\pd[3]\deltae\right)+\realvei[2]\left(\imvei[1]\pd[3]\deltae-\imvei[3]\pd[1]\deltae\right)+\realvei[3]\left(\imvei[2]\pd[1]\deltae-\imvei[1]\pd[2]\deltae\right)\right]^{2}},\label{eq:tdoe-2}
\end{equation}
and
\begin{equation}
\left\langle \delta^{2}\widehat{\alpha}_{3}\right\rangle =\frac{t^{2}\csc^{2}\left(\frac{1}{2}\deltae t\right)\left[\left(\realvei[1]\pd[2]\deltae-\realvei[2]\pd[1]\deltae\right)^{2}+\left(\imvei[1]\pd[2]\deltae-\imvei[2]\pd[1]\deltae\right)^{2}\right]+16\left(\realvei[2]\imvei[1]-\realvei[1]\imvei[2]\right)^{2}}{16nt^{2}\left[\realvei[1]\left(\imvei[3]\pd[2]\deltae-\imvei[2]\pd[3]\deltae\right)+\realvei[2]\left(\imvei[1]\pd[3]\deltae-\imvei[3]\pd[1]\deltae\right)+\realvei[3]\left(\imvei[2]\pd[1]\deltae-\imvei[1]\pd[2]\deltae\right)\right]^{2}},\label{eq:tdoe-3}
\end{equation}

\noindent respectively, where
\begin{equation}
\begin{array}{rl}
\deltae= & \eigen[0]-\eigen[1],\\
\realvei[i]= & \mathrm{Re}\left(\left\langle \eigen[0]\right|\left.\partial_{i}\eigen[1]\right\rangle \right),\\
\imvei[i]= & \mathrm{Im}\left(\left\langle \eigen[0]\right|\left.\partial_{i}\eigen[1]\right\rangle \right).
\end{array}\label{eq:tdoe-4}
\end{equation}

Since $\left\langle \delta^{2}\widehat{\alpha}_{1}\right\rangle $,
$\left\langle \delta^{2}\widehat{\alpha}_{2}\right\rangle $, and
$\left\langle \delta^{2}\widehat{\alpha}_{3}\right\rangle $ are symmetric
with respect to each other, we consider only $\left\langle \delta^{2}\widehat{\alpha}_{1}\right\rangle $
in the following analysis. When $\left(\realvei[2]\pd[3]\deltae-\realvei[3]\pd[2]\deltae\right)^{2}+\left(\imvei[2]\pd[3]\deltae-\imvei[3]\pd[2]\deltae\right)^{2}=0$,
the oscillatory term vanishes. To ensure a nonzero numerator, we have
$\pd[2]\deltae=\pd[3]\deltae=0$, leading to
\begin{equation}
\left\langle \delta^{2}\widehat{\alpha}_{1}\right\rangle =\frac{1}{nt^{2}\left(\pd[1]\deltae\right)^{2}},\label{eq:tdoe-5}
\end{equation}
where $\pd[1]\deltae\neq0$. In this case, $\left\langle \delta^{2}\widehat{\alpha}_{1}\right\rangle $
achieves the Heisenberg limit. For instance, consider a Hamiltonian
\begin{equation}
H_{\left(B,\theta,\varphi\right)}=B\left(\cos\left(\theta\right)\cos\left(\varphi\right)\sigma_{x}+\cos\left(\theta\right)\sin\left(\varphi\right)\sigma_{y}+\sin\left(\theta\right)\sigma_{z}\right),\label{eq:tdoe-6}
\end{equation}
where $\sigma_{x}$, $\sigma_{y}$, and $\sigma_{z}$ are Pauli matrices.
For estimating $B$, the fact that the eigenvalues of the Hamiltonian
are independent of $\theta$ and $\varphi$ enables the estimation
precision to reach the Heisenberg limit.

When $\left(\realvei[2]\pd[3]\deltae-\realvei[3]\pd[2]\deltae\right)^{2}+\left(\imvei[2]\pd[3]\deltae-\imvei[3]\pd[2]\deltae\right)^{2}\neq0$,
we first consider two extreme cases. When $t$ is sufficiently small
such that $t\ll1/\left|\deltae\right|$, we have $\csc^{2}\left(\frac{1}{2}\deltae t\right)\approx\frac{4}{\left(\deltae t\right)^{2}}$,
so
\begin{equation}
\left\langle \delta^{2}\widehat{\alpha}_{1}\right\rangle \propto\frac{1}{t^{2}}.\label{eq:tdoe-7}
\end{equation}
When $t$ is sufficiently large, we have
\begin{equation}
\left\langle \delta^{2}\widehat{\alpha}_{1}\right\rangle \propto\csc^{2}\left(\frac{1}{2}\deltae t\right),\label{eq:tdoe-8}
\end{equation}
which is a periodic function. For general $t$, its overall trend
can be analyzed via periodic sampling. We set the initial sampling
time as $\ts[0]\in\left(0,2\pi/\left|\deltae\right|\right)$, and
define $c_{0}=\csc^{2}\left(\frac{1}{2}\deltae\ts[0]\right)$. For
$\ts[k]=\ts[0]+\frac{2k\pi}{\left|\deltae\right|}$ with $k=1,2,3,\ldots$
, we have $\csc^{2}\left(\frac{1}{2}\deltae\ts[k]\right)=c_{0}$.
The function
\begin{equation}
\left\langle \delta^{2}\widehat{\alpha}_{1}\right\rangle _{c_{0}}=\frac{c_{0}t^{2}\xi_{1}+\xi_{2}}{nt^{2}\xi_{3}},\label{eq:tdoe-9}
\end{equation}
passes through all these sampling points, where
\begin{equation}
\begin{array}{rl}
\xi_{1}= & \left(\realvei[2]\pd[3]\deltae-\realvei[3]\pd[2]\deltae\right)^{2}+\left(\imvei[2]\pd[3]\deltae-\imvei[3]\pd[2]\deltae\right)^{2},\\
\xi_{2}= & 16\left(\realvei[3]\imvei[2]-\realvei[2]\imvei[3]\right)^{2},\\
\xi_{3}= & 16\left[\realvei[1]\left(\imvei[3]\pd[2]\deltae-\imvei[2]\pd[3]\deltae\right)+\realvei[2]\left(\imvei[1]\pd[3]\deltae-\imvei[3]\pd[1]\deltae\right)+\realvei[3]\left(\imvei[2]\pd[1]\deltae-\imvei[1]\pd[2]\deltae\right)\right]^{2}.
\end{array}\label{eq:tdoe-10}
\end{equation}
Since
\begin{equation}
\pd[t]\left\langle \delta^{2}\widehat{\alpha}_{1}\right\rangle _{c_{0}}\leq0,\label{eq:tdoe-11}
\end{equation}
the variance shows an overall decreasing trend. From Eq.~\eqref{eq:tdoe-8},
we know that at large $t$, the minimum variance per period occurs
at $t_{k}=\left(2k+1\right)\pi/\left|\deltae\right|$, i.e., $c_{0}=1$.
Therefore, the infimum of the variance is
\begin{equation}
\inf\left\langle \delta^{2}\widehat{\alpha}_{1}\right\rangle =\frac{\xi_{1}}{n\xi_{3}}.\label{eq:tdoe-12}
\end{equation}
However, since $\left|\pd[t]\left\langle \delta^{2}\widehat{\alpha}_{1}\right\rangle _{c_{0}}\right|$
decays cubically with time, it is not worthwhile to expend more time
for limited gains in precision.

\refstepcounter{suppnote}

\section*{Supplementary Note 6. Necessity for reparameterization}\label{sec:note6}

This Supplementary Note shows the necessity for reparameterization.
In general, the parameterization of the Hamiltonian for a two-dimensional
system can be arbitrary. For example, one can parameterize the Hamiltonian
by the coefficients in the Pauli basis or by the strength and angular
parameters. For the current problem of reducing $\deltaes$ so as
to extend the evolution time for the Heisenberg scaling, it will turn
out that the parameterization by the coefficients in the Pauli basis
is more convenient, so we will focus on this parameterization.

Usually the unknown parameters can have complex correlations between
each other, manifested by the correlation matrix of the estimation.
To simplify the estimation task of these parameters, a powerful tool
is to transform the unknown parameters to other parameters which have
simpler correlations (e.g., a diagonal correlation matrix) \citep{suzuki2020quantum,albarelli2020aperspective},
sometimes known as reparametrization. If the vector of the parameters
$\boldsymbol{\alpha}$ is reparameterized as $\boldsymbol{\beta}=\left(\beta_{1},\beta_{2},\ldots,\beta_{k}\right)$,
the quantum Fisher information matrix of $\boldsymbol{\beta}$ is
related to the original quantum Fisher information matrix for $\boldsymbol{\alpha}$
through
\begin{equation}
F_{\boldsymbol{\beta}}=J^{\mathrm{T}}F_{\boldsymbol{\alpha}}J,\label{eq:nfr-1}
\end{equation}
where $J$ is the Jacobian matrix defined as $J_{ij}=\partial\alpha_{i}/\partial\beta_{j}$.

Suppose the original Hamiltonian of the probe is decomposed in the
Pauli basis as
\begin{equation}
\hainit=\fav[\boldsymbol{\alpha}]\cdot\boldsymbol{\sigma},\label{eq:nfr-2}
\end{equation}
where $\boldsymbol{\sigma}=\left(\sigma_{x},\sigma_{y},\sigma_{z}\right)$,
$\boldsymbol{\alpha}=\left(\alpha_{1},\alpha_{2},\alpha_{3}\right)$,
and $\fav[\boldsymbol{\alpha}]=\left(\fa[1][\boldsymbol{\alpha}],\fa[2][\boldsymbol{\alpha}],\fa[3][\boldsymbol{\alpha}]\right)$
with $\fa[1][\boldsymbol{\alpha}]$, $\fa[2][\boldsymbol{\alpha}]$,
and $\fa[3][\boldsymbol{\alpha}]$ being real-valued functions. The
control Hamiltonian at the $\left(k+1\right)$-th iteration is denoted
by
\begin{equation}
\hcontkth[k+1]=-\fav[\hat{\boldsymbol{\alpha}}_{k}]\cdot\boldsymbol{\sigma},\label{eq:nfr-3}
\end{equation}
where $\hat{\boldsymbol{\alpha}}_{k}=\left(\hat{\alpha}_{k,1},\hat{\alpha}_{k,2},\hat{\alpha}_{k,3}\right)$
denotes the control parameters which are essentially the estimates
of $\hat{\boldsymbol{\alpha}}$ from the $k$-th iteration. Suppose
$\hat{\boldsymbol{\alpha}}_{k}$ deviates from the true value of $\boldsymbol{\alpha}$
by $\deltaak[k]$, i.e.,
\begin{equation}
\hat{\boldsymbol{\alpha}}_{k}=\boldsymbol{\alpha}_{0}+\deltaak[k],\label{eq:nfr-4}
\end{equation}
with $\boldsymbol{\alpha}_{0}=\left(\alpha_{0,1},\alpha_{0,2},\alpha_{0,3}\right)$
being the true value of $\boldsymbol{\alpha}$ and $\deltaak[k]=\left(\delta\alpha_{k,1},\delta\alpha_{k,2},\delta\alpha_{k,3}\right)$
being the estimation errors from the $k$-th iteration, we obtain
the $\deltaes$ of the $\left(k+1\right)$-th iteration as
\begin{equation}
\deltaesk[k+1]=4\left\Vert \fav[\boldsymbol{\alpha}_{0}]-\fav[\hat{\boldsymbol{\alpha}}_{k}]\right\Vert ^{2}.\label{eq:nfr-5}
\end{equation}
When $\left\Vert \deltaak[k]\right\Vert \ll1$, Eq.~\eqref{eq:nfr-5}
is simplified to
\begin{equation}
\deltaesk[k+1]\approx4\left\Vert J\deltaak[k]\right\Vert ^{2},\label{eq:nfr-6}
\end{equation}
where $J$ is the Jacobian matrix defined by $J_{ij}=\partial_{\alpha_{j}}\fa[i][\boldsymbol{\alpha}]$
evaluated at $\boldsymbol{\alpha}_{0}$.

If we reparameterize the Hamiltonian as
\begin{equation}
\hainit[\boldsymbol{\beta}]=\beta_{1}\sigma_{x}+\beta_{2}\sigma_{y}+\beta_{3}\sigma_{z},\label{eq:nfr-7}
\end{equation}
where the new parameters are defined as $\beta_{1}=\fa[1][\boldsymbol{\alpha}]$,
$\beta_{2}=\fa[2][\boldsymbol{\alpha}]$, and $\beta_{3}=\fa[3][\boldsymbol{\alpha}]$.
According to Eq.~\eqref{eq:nfr-5}, we obtain
\begin{equation}
\deltaesk[k+1]=4\left\Vert \deltabk[k]\right\Vert ^{2},\label{eq:nfr-8}
\end{equation}
which will greatly simplify the subsequent optimization.

\section*{Supplementary Note 7. Derivation of covariance matrix}\label{sec:note7}

In this Supplementary Note, we derive the covariance matrix for the
$k$-th iteration.

Suppose the estimate from the $\left(k-1\right)$-th iteration is
$\hat{\boldsymbol{\beta}}_{k-1}$. The control Hamiltonian in the
$k$-th iteration is $\hcontkth[k]=\hat{\boldsymbol{\beta}}_{k-1}\cdot\boldsymbol{\sigma}$
and the total Hamiltonian is $\ha[\boldsymbol{\beta},k]=\left(\boldsymbol{\beta}-\hat{\boldsymbol{\beta}}_{k-1}\right)\cdot\boldsymbol{\sigma}$.
From Eq.~(10) in the main manuscript, the entries of quantum Fisher
information matrix are
\begin{equation}
\begin{array}{rl}
F_{k,ii}^{\left(\boldsymbol{\beta}\right)}= & \frac{4\left(\deltabki[k-1][i]^{2}\left\Vert \deltabk[k-1]\right\Vert ^{2}t_{k}^{2}+\left(\left\Vert \deltabk[k-1]\right\Vert ^{2}-\deltabki[k-1][i]^{2}\right)\sin^{2}\left(\left\Vert \deltabk[k-1]\right\Vert t_{k}\right)\right)}{\left\Vert \deltabk[k-1]\right\Vert ^{4}},\\
F_{k,ij}^{\left(\boldsymbol{\beta}\right)}= & \frac{4\deltabki[k-1][i]\deltabki[k-1][j]\left(\left\Vert \deltabk[k-1]\right\Vert ^{2}t_{k}^{2}-\sin^{2}\left(\left\Vert \deltabk[k-1]\right\Vert t_{k}\right)\right)}{\left\Vert \deltabk[k-1]\right\Vert ^{4}},i\neq j.
\end{array}\label{eq:doc-1}
\end{equation}
The covariance matrix in the $k$-th iteration is given by
\begin{equation}
\covkb[k]=\left(n_{k}F_{k}^{\left(\boldsymbol{\beta}\right)}\right)^{-1},\label{eq:doc-2}
\end{equation}
the elements of which turn out to be
\begin{equation}
\begin{array}{rl}
\covkb[k,ii]= & \frac{\frac{\deltabki[k-1][i]^{2}}{t_{k}^{2}\left\Vert \deltabk[k-1]\right\Vert ^{2}}+\left[\left\Vert \deltabk[k-1]\right\Vert ^{2}-\deltabki[k-1][i]^{2}\right]\csc^{2}\left(\left\Vert \deltabk[k-1]\right\Vert t_{k}\right)}{4n_{k}},\\
\covkb[k,ij]= & \frac{\deltabki[k-1][i]\deltabki[k-1][j]\left(\frac{1}{t_{k}^{2}\left\Vert \deltabk[k-1]\right\Vert ^{2}}-\csc^{2}\left(\left\Vert \deltabk[k-1]\right\Vert t_{k}\right)\right)}{4n_{k}},i\neq j.
\end{array}\label{eq:doc-3}
\end{equation}

\refstepcounter{suppnote}

\section*{Supplementary Note 8. Relation between estimation variances of the
original parameter with adaptive and optimal control}\label{sec:note8}

In this Supplementary Note, we derive the relation between estimation
variances of the original parameter with adaptive and optimal control
strategy.

Suppose the estimation variance obtained from the adaptive control
strategy with $m$ iterations is $\vsum[m]=\mathrm{Tr}\covkb[m]$,
and that from the optimal control strategy with the true values of
the unknown parameters is $\vsum[\mathrm{oc}]$, both for the parameters
in the Pauli basis, $\hat{\boldsymbol{\beta}}_{1}$, $\hat{\boldsymbol{\beta}}_{2}$,
and $\hat{\boldsymbol{\beta}}_{3}$ after reparameterization, rather
than the original parameters in the Hamiltonian. If the estimation
variance of the original parameter $\alpha_{i}$ using the adaptive
control strategy approaches the variance obtained with the optimal
control strategy, analogous to how $\vsum[m]$ approaches $\vsum[\mathrm{oc}]$,
it means the effectiveness of our adaptive control strategy in estimating
the original parameters.

For the adaptive control strategy, the covariance matrix after $m$
iterations is $\covkb[m]$, with elements given in Eq.~\eqref{eq:doc-3}.
When the optimal evolution time scheme is applied, the total variance
is
\begin{equation}
\vsum[m]=\frac{\frac{\left\Vert \deltabk[m-1]\right\Vert ^{2}}{\prodz^{2}}+2\left\Vert \deltabk[m-1]\right\Vert ^{2}\csc^{2}\left(\prodz\right)}{4\tnk[m]},\label{eq:rbe-1}
\end{equation}
where we have used Eq.~(25) in the main manuscript. For the optimal
control Hamiltonian, $\vsum[\mathrm{oc}]$ is given by Eq.~(33) in
the main manuscript.

When the Hamiltonian contains three independent unknown parameters,
$\boldsymbol{\alpha}=\left(\alpha_{1},\alpha_{2},\alpha_{3}\right)$,
the quantum Fisher information matrix of $\boldsymbol{\alpha}$ is
related to the quantum Fisher information matrix of $\boldsymbol{\beta}$
via
\begin{equation}
F_{\boldsymbol{\alpha}}=J^{\mathrm{T}}F_{\boldsymbol{\beta}}J,\label{eq:rbe-2}
\end{equation}
where $J$ is given by $J_{ij}=\partial\beta_{i}/\partial\alpha_{j}$,
with $\beta_{i}=\fa[i][\boldsymbol{\alpha}]$. Therefore, the relation
between their covariance matrices is
\begin{equation}
\covka=J^{-1}\covkb\left(J^{-1}\right)^{\mathrm{T}}.\label{eq:rbe-3}
\end{equation}

For the adaptive control strategy, Eq.~\eqref{eq:rbe-3} yields the
variances of the original parameters after $m$ iterations as
\begin{equation}
\begin{array}{rl}
\covka[m,ii]= & \frac{1}{4\tnk[m]}\sum_{r,s=1,r\neq s}^{3}\left(J^{-1}\right)_{ir}\left(J^{-1}\right)_{is}\deltabki[m-1][r]\deltabki[m-1][s]\left(\frac{1}{\prodzs}-\csc^{2}\left(\prodz\right)\right)\\
 & +\frac{1}{4\tnk[m]}\sum_{r=1}^{3}\left(J^{-1}\right)_{ir}^{2}\left[\frac{\deltabki[m-1][r]^{2}}{\prodzs}+\left(\left\Vert \deltabk[m-1]\right\Vert ^{2}-\deltabki[m-1][r]^{2}\right)\csc^{2}\left(\prodz\right)\right],
\end{array}\label{eq:rbe-4}
\end{equation}
where we have used the optimal evolution time scheme. Let
\begin{equation}
\begin{array}{rl}
\mu_{\mathrm{max},i}= & \max\left\{ \left.\left|\left(J^{-1}\right)_{ir}\left(J^{-1}\right)_{is}\right|\ \right|r,s\in\left\{ 1,2,3\right\} ,r\neq s\right\} ,\\
\nu_{\mathrm{max},i}= & \max\left\{ \left.\left(J^{-1}\right)_{ir}^{2}\right|r\in\left\{ 1,2,3\right\} \right\} .
\end{array}\label{eq:rbe-5}
\end{equation}
We obtain
\begin{equation}
\covka[m,ii]\leq\left(2\mu_{\mathrm{max},i}\frac{\prodzs\csc^{2}\left(\prodz\right)-1}{1+2\prodzs\csc^{2}\left(\prodz\right)}+\nu_{\mathrm{max},i}\right)\vsum[m].\label{eq:rbe-6}
\end{equation}
For the optimal control strategy, Eq.~\eqref{eq:rbe-3} yields the
variances of the original parameters as
\begin{equation}
\begin{array}{rl}
\covka[\mathrm{oc},ii]= & \frac{\left(J^{-1}\right)_{i1}^{2}+\left(J^{-1}\right)_{i2}^{2}+\left(J^{-1}\right)_{i3}^{2}}{4\tnoc\tkoc^{2}}\\
\geq & \frac{\nu_{\mathrm{max},i}}{3}\vsum[\mathrm{oc}].
\end{array}\label{eq:rbe-7}
\end{equation}
Using
\begin{equation}
\vsum[m]=\gap\vsum[\mathrm{oc}]\label{eq:rbe-8}
\end{equation}
to connect Eq.~\eqref{eq:rbe-6} and Eq.~\eqref{eq:rbe-7}, we obtain
\begin{equation}
\covka[m,ii]\leq\frac{12\prodzs\csc^{2}\left(\prodz\right)-3}{1+2\prodzs\csc^{2}\left(\prodz\right)}\gap\covka[\mathrm{oc},ii],\label{eq:rbe-9}
\end{equation}
where $\mu_{\mathrm{max},i}\leq\nu_{\mathrm{max},i}$ has been used.

\refstepcounter{suppnote}

\section*{Supplementary Note 9. Probability density function of deviation factor}\label{sec:note9}

In this Supplementary Note, we derive the probability density function
of deviation factor $\dev$ for the $k$-th iteration.

Since $\deltaesk[k]=4\left\Vert \deltabk[k-1]\right\Vert ^{2}$ and
$\deltabk[k-1]$ follows a Gaussian distribution $\deltabk[k-1]\sim\mathcal{N}\left(\boldsymbol{0},\covkb[k-1]\right)$
in the asymptotic limit of the number of trials, $\deltaesk[k]$ follows
a generalized $\chi^{2}$ distribution, expressed as a weighted sum
of squares of independent standard normal random variables, where
the weights are four times the eigenvalues of $\covkb[k-1]$. Together
with $\left\Vert \deltabk[k-2]\right\Vert \toptk[][k-1]=\prodz$,
we obtain
\begin{equation}
\deltaesk[k]=\frac{1}{\tnk[k-1]\toptk[][k-1]^{2}}\chi_{1}^{2}\left(1\right)+\frac{\prodzs\csc^{2}\left(\prodz\right)}{\tnk[k-1]\toptk[][k-1]^{2}}\chi_{2}^{2}\left(1\right)+\frac{\prodzs\csc^{2}\left(\prodz\right)}{\tnk[k-1]\toptk[][k-1]^{2}}\chi_{3}^{2}\left(1\right).\label{eq:dod-1}
\end{equation}
The mean of $\deltaesk[k]$ is $\left(2\prodzs\csc^{2}\left(\prodz\right)+1\right)/\left(\tnk[k-1]\toptk[][k-1]^{2}\right)$.
From Eq.~(35) of the main manuscript, we have
\begin{equation}
\dev=\frac{\chi_{1}^{2}\left(1\right)+\prodzs\csc^{2}\left(\prodz\right)\left(\chi_{2}^{2}\left(1\right)+\chi_{3}^{2}\left(1\right)\right)}{2\prodzs\csc^{2}\left(\prodz\right)+1}.\label{eq:dod-2}
\end{equation}
Using the Laplace transform, we obtain the probability density function
of $\dev$ as
\begin{equation}
f_{\dev}=\frac{\left(2\prodz^{2}+\sin^{2}\left(\prodz\right)\right)\mathrm{e}^{-\frac{\dev\sin^{2}\left(\prodz\right)}{2\prodz^{2}}-\dev}\text{erf}\left(\frac{\sqrt{\dev\left(-\frac{\sin^{2}\left(\prodz\right)}{\prodz^{2}}+2\prodz^{2}\csc^{2}\left(\prodz\right)-1\right)}}{\sqrt{2}}\right)}{2\prodz\sqrt{\prodz^{2}-\sin^{2}\left(\prodz\right)}},\label{eq:dod-3}
\end{equation}
where
\begin{equation}
\text{erf}\left(x\right)=\frac{2}{\sqrt{\pi}}\int_{0}^{x}\mathrm{e}^{-t^{2}}dt.\label{eq:dod-4}
\end{equation}

\refstepcounter{suppnote}
\end{document}